\documentclass[epj,fleqn]{svjour}
\usepackage{amsmath}
\usepackage{amssymb}
\usepackage{epsfig}
\numberwithin{equation}{section}

\def\ep{{\epsilon}}

\def\k{{{\bf k}}}
\def\om{{\omega}}

\def\q{{{\bf q}}}

\def\e{{$\grave{e}$}}
\def\nnu{{\nonumber}}

\def\g{{\bf{g}}}

\def\q{{\bf{q}}}

\def\beq{\begin{equation}}
\def\eeq{\end{equation}}
\def\beqa{\begin{eqnarray}}
\def\eeqa{\end{eqnarray}}

\def\eft{{\tilde{\epsilon}}_f}

\def\g0{{\gamma_0}}
\def\ii{{\mbox{i}}}
\def\sgn{{\mbox{sgn}}}

\def\Im{{\mbox{Im}}}
\def\H{{\mbox{H}}}

\def\dna{{\downarrow}}
\def\upa{{\uparrow}}
\def\mj{M. Jarrell}
\def\tp{T. Pruschke}
\def\del{D. E. Logan}
\def\rb{R. Bulla}
\def\prb{Phys. Rev. B }
\def\prl{Phys. Rev. Lett. }
\def\jpcm{J. Phys. Condens. Matt. }

\def\rmp{Rev. Mod. Phys. }
\def\nsv{N. S. Vidhyadhiraja}
\def\hrk{H. R. Krishnamurthy}
\begin{document}

\title{Dynamics and scaling in the periodic Anderson model}
\titlerunning{Dynamics and scaling in the periodic Anderson model}
\authorrunning{}
\author{N.S.Vidhyadhiraja and David.E.Logan
}                     
%
%
\institute{Oxford University, Physical and Theoretical Chemistry Laboratory,
South Parks Road, Oxford OX1 3QZ, UK. }

\date{Received: date / Revised version: date}
\abstract{
  The periodic Anderson model (PAM) captures the essential
physics of heavy fermion materials. Yet even for the paramagnetic metallic
phase, a practicable many-body theory that can simultaneously handle all energy scales
while respecting the dictates of Fermi liquid theory at low energies, 
and all interaction strengths from the strongly correlated Kondo lattice through to 
weak coupling, has remained quite elusive.
Aspects of this problem are considered in the present paper
where a non-perturbative local moment approach (LMA) to single-particle dynamics of
the asymmetric PAM is developed within
the general framework of dynamical mean-field theory.
All interaction strengths and energy scales are encompassed, although
our natural focus is the Kondo lattice regime of essentially
localized $f$-spins but general conduction band filling, characterised by
an exponentially small lattice coherence scale $\omega_{L}$. Particular
emphasis is given to the resultant universal scaling behaviour of dynamics
in the Kondo lattice regime
as an entire function of $\omega^{\prime} =\omega/\omega_{L}$, including 
its dependence on conduction band filling, $f$-level asymmetry and lattice type.
A rich description arises, 
encompassing both coherent Fermi liquid behaviour at low-$\omega^{\prime}$ 
and the crossover to effective single-impurity scaling physics at higher
energies --- but still in the $\omega/\omega_{L}$-scaling regime, and
as such incompatible with the presence of two-scale `exhaustion' 
physics, which is likewise discussed.
} 

\maketitle

\section{Introduction.}
\label{sec:intro}

  The paramagnetic metallic phase of heavy fermion materials
provides a classic example of strongly correlated electron physics [1,2].
Spin-flip scattering of itinerant conduction electrons by essentially localized
$f$-level electrons leads to large effective masses and the low-energy
scale(s) symptomatic of any strongly correlated state. 
At low energies and/or temperatures the lattice coherence is paramount
and the system is a Fermi liquid with well defined 
quasiparticles and coherent screening of the $f$-level spins; behaviour that
crosses over for sufficiently high energies to essentially incoherent screening and 
the effective single-impurity characteristics of the Kondo effect [1,2].
 
Handling theoretically the many sides and attendant issues of this basic physics
is of course another matter. The paradigm here is the periodic Anderson model (PAM), 
the natural lattice generalization of the Anderson impurity model (AIM), in which each 
lattice site contains a correlated $f$-level (with interaction $U$) hybridizing 
locally with a non-interacting conduction band [1,2]; and a description of 
which remains a major challenge,
particularly in the strongly correlated Kondo lattice regime of effectively 
localized $f$-spins but arbitrary conduction
band filling. That reflects in large part the inherent difficulties in developing
a many-body theory that can capture non-perturbatively
the strong coupling regime of primary interest, satisfying in particular the dictates
of Fermi liquid theory at low energies and yet capable of describing the problem on
all energy scales. Moreover, no matter how strong the
correlations, the Fermi liquid nature of the ground state implies adiabatic
continuity to the non-interacting limit; so the same theory should also be able to
handle the full range of interaction strengths, including simple
perturbative behaviour in weak coupling.

  Our aim in the present paper is to develop an approach to the paramagnetic
phase of the PAM that meets the above criteria, within the general framework
of dynamical mean-field theory (DMFT) 
[3-6]. The PAM has of course been studied
extensively within DMFT using an impressive range of 
methods. Numerical techniques include the numerical renormalization group
(NRG) [7,8] quantum Monte Carlo (QMC) [9-12] and exact diagonalization [13]. Theoretical
approaches range from perturbation theory in the interaction $U$ [14,15], iterated
perturbation theory [16,17], the lattice non-crossing approximation [18,19] and the
average $t$-matrix approximation [20], to large-$N$ mean-field theory/slave bosons
[21-23] and the Gutzwiller variational approach [24,25]. 
Such techniques nonetheless possess well known limitations [2]; be it an
inability to handle strong correlations, failure to recover Fermi liquid
behaviour or even the non-interacting limit, unrealistic confinement to the lowest
energy scales and so on. 
NRG aside, analogous comments apply to full scale numerical 
methods. QMC for example is restricted to modest interactions and relatively high
temperatures, while finite-size effects render exact diagonalization of limited value.
These remarks are certainly not intended to detract from the many insights that
have accrued from such approaches. They are made simply to emphasise the 
desirability of developing new, necessarily approximate theories for this
longstanding problem.

  One such is pursued here, the local moment approach (LMA) [26-35], the primary
emphasis of which is on single-particle dynamics and transport. Initially  
developed in the context of pure quantum impurity models (AIMs) [26-33], the LMA is
intrinsically non-perturbative but technically quite simple, with the 
physically intuitive notion of local moments introduced explicitly from the outset.
This leads naturally to an underlying `two-self-energy' description
in which the essential correlated spin-flip physics is
readily captured; and corresponds physically to dynamical tunneling between
initially degenerate local moment configurations, which in lifting
the spin-degeneracy restores the local singlet symmetry characteristic of the Fermi
liquid state.
The desiderata mentioned above are well met [26-33], all interaction strengths
and energy scales being encompassed, including the low-energy requirements of Fermi 
liquid theory (although the approach can also handle models with non-Fermi liquid 
phases, see e.g. [31-33]). In particular, for the strong coupling Kondo regime of the 
conventional metallic AIM, LMA results for dynamics have been shown [28,29,33] to 
give very good agreement with  NRG calculations; and, for static magnetic 
properties, with exact results from the Bethe ansatz [30].

More recently, exploiting the fact that within DMFT all correlated lattice-fermion 
models reduce to an effective quantum impurity hybridizing self-consistently with 
the surrounding fermionic bath [3-6], the LMA has been extended
to encompass the particle-hole symmetric PAM [34,35]. Here 
the system is ubiquitously a `Fermi liquid insulator' that
evolves continuously with increasing interaction strength
from a simple non-interacting hybridization-gap insulator to the strongly
correlated Kondo insulating state; with an insulating gap scale 
that becomes exponentially small in strong coupling, such that physical
properties exhibit universal scaling in terms of it ({\it i.e.\ }contain \it no \rm
explicit dependence on the `bare' high-energy material parameters, $U$ etc, that
enter the underlying model Hamiltonian).
A comprehensive description of single-particle dynamics [34,35], electrical transport
and optical properties [35] of Kondo insulators arises, encompassing
all relevant frequency ($\om$) and/or temperature ($T$) domains; and
exploitation of scaling in particular enables direct, rather successful comparison to 
a range of experiments [35].

  Important though it is to the problem of Kondo insulators the particle-hole 
symmetric PAM is of course special, and the desirability of developing the
LMA to encompass the asymmetric PAM and hence the generic case of heavy fermion 
metals is self-evident. That is considered here, our specific focus being
on $T=0$ single-particle dynamics.
In addition to intrinsic interest in such 
{\it per se}, and the fact that 
their $\om$-dependence exemplifies much of the underlying physics of the problem, 
knowledge of single-particle dynamics is well known [3-6] to be sufficient within 
DMFT to determine $\q =0$ transport and optical properties; 
which will be considered
in a subsequent paper (in parallel to
previous LMA work on Kondo insulators [34,35]).
The present paper is accordingly organised as follows. 
After appropriate background to the PAM within DMFT (\S 2), formulated
for an essentially arbitrary lattice,
implications of adiabatic continuity and the Luttinger integral theorem [36] are
considered in \S 3; together with the quasiparticle forms for the local conduction
($c$-) and $f$-electron spectra that Fermi liquid theory requires be satisfied on
the lowest energies $|\om| \lesssim \om_L$, where $\om_L$ is the
low-energy scale characteristic of the coherent Fermi liquid state. The LMA
itself is considered in \S 4, first in general terms applicable to an
essentially arbitrary diagrammatic approximation for the inherent
two-self-energies, and including the issue of 
symmetry restoration that is central to the approach. The specific non-perturbative
approximation to the LMA self-energies implemented here is then discussed, 
together with the practical method of solution such that the dictates of both symmetry
restoration and the Luttinger theorem are satisfied.

  Results arising are presented in \S 5, with a natural 
emphasis on the strongly correlated Kondo lattice regime.
An overview of dynamics on all energy scales is first given (\S 5.1), 
encompassing both the `low'-energy behaviour characteristic of the 
renormalized heavy electron state as well as non-universal energies 
on the order of bare bandwidths or the interaction $U$. In addition to 
illustrating the broad roles of asymmetry (in both the conduction band 
and $f$-levels), and of lattice type, comparison is 
also made on this `all scales' level both to results for single-particle 
dynamics of the AIM (in which only a single correlated $f$-level 
is coupled to the conduction band), and to dynamics of the PAM arising 
at the crude level of static mean-field. In \S 5.2 the material 
dependence of the low-energy
lattice coherence scale $\om_L$ on bare model parameters is obtained, in
the strong coupling Kondo lattice regime where $\om_L$ is exponentially small; 
and its behaviour compared in turn to corresponding LMA results for the 
AIM Kondo 
scale $\om_K$. The central issues of scaling are considered in \S 5.3: the 
resultant universal scaling behaviour of dynamics in terms of 
$\om^{\prime} = \om/\om_L$, on {\it all } $\om^{\prime}$ scales,
and including the dependence of scaling dynamics on conduction band filling. 
At low-$\om^{\prime}$ the scaling spectra exhibit coherent Fermi liquid behaviour,
crossing over with increasing energy to logarithmically slow spectral tails.
The latter are found to be independent of both conduction band filling and
lattice type, and to
have precisely the same {\it scaling form } as those for an AIM; 
establishing thereby the crossover from low-energy coherent Fermi liquid
behaviour to effective incoherent single-impurity physics on high-$\om^{\prime}$
scales, but still in the $\om/\om_L$-scaling regime. A discussion of 
results obtained here in relation to the issue of two-scale `exhaustion' physics
[37,38] is given in \S 5.4; and some concluding remarks are made in \S 6.

\section{Background}
\label{sec:model}

  The Hamiltonian for the PAM, $\hat{H} = \hat{H}_c + \hat{H}_f
  +\hat{H}_{hyb}$, is given in standard notation by:
\beq
\begin{split}
&\hat{H}=\sum_{i,\sigma}\ep_c c^\dag_{i\sigma} 
c^{\phantom{\dag}}_{i\sigma}  
-t\sum_{(i,j),\sigma} c^\dag_{i\sigma} c^{\phantom{\dag}}_{j\sigma}\\
&+ \sum_{i,\sigma}\left(\ep_f 
+ 
\tfrac{U}{2}  
f^\dag_{i\,-\sigma}f^{\phantom{\dagger}}_{i\,-\sigma}\right)
f^\dag_{i\sigma}f^{\phantom{\dag}}_{i\sigma}  \label{eq:model}  
+V\sum_{i,\sigma} (f^\dag_{i\sigma} c^{\phantom{\dag}}_{i\sigma} +\mbox{h.c.})
\end{split}
\eeq
The first two terms describe the uncorrelated conduction ($c$) band,
$\hat{H}_c$; with $c$-orbital site energies $\ep_c$ and 
nearest neighbour hoppings $t_{ij}=t$, rescaled as 
$t\propto t_*/\surd Z_c$
in the large dimensional limit where the coordination number 
$Z_c \rightarrow \infty$ [3-6]
(with $t_*$ the basic energy unit).
 The second term, $\hat{H}_f$, describes  the correlated $f$-levels,
with site energies $\ep_f$ and on-site repulsion $U_{ff}=U$; while 
the final term, $\hat{H}_{hyb}$ couples the $c$- and $f$- subsystems
via the local hybridization matrix element $V$. Throughout the paper
the Fermi level is taken as the origin of energy, $\om_F=0$.

The model is thus 
characterized by four independent `bare'/material  parameters, 
namely $\ep_c/t_*, V/t_*,  \ep_f/t_*$ and $U/t_*$ (with $t_*\equiv 1$
taken throughout) -- a huge parameter space in comparison e.g.\ to
the Hubbard model. In previous LMA work on the PAM [34,35] we have studied 
the particle-hole symmetric model appropriate to the Kondo insulating
state; for which $\ep_f=-\tfrac{U}{2}$ and $\ep_c=0$, with consequent
occupancies $n_f\!\!=\sum_\sigma\!\!<\!\!f^\dag_{i\sigma}f^{\phantom{\dag}}_{i\sigma}\!\!>\,=\!1$
and 
$n_c\!\!=\sum_\sigma\!\!<\!\!c^\dag_{i\sigma}c^{\phantom{\dag}}_{i\sigma}\!\!>\,=\!1$
for all $U$. Here we consider the generic asymmetric case, 
encompassing heavy Fermion metals (and with the symmetric PAM
recovered as a particular limit). Particle-hole 
asymmetry itself enters the problem in two ways [8] . (a) Conduction band
asymmetry, reflected in $\ep_c\neq 0$ which, as detailed below, specifies the 
centre of gravity of the free ($V=0$) conduction band relative to the Fermi
level. (b) $f$-level asymmetry which, as for an impurity Anderson
model [27], is embodied in the parameter 
\beq
\eta =  1 +\frac{2\ep_f}{U}
\eeq
such that $\eta=0$ corresponds to particle-hole symmetric $f$-levels.
The bare parameter set may thus be taken equivalently as $\ep_c, V, U$ and
$\eta$. We shall find this choice to be convenient in the following
(and in fact necessary to describe universal scaling behaviour in the Kondo 
lattice regime, see \S5.3).

While the LMA developed here encompasses all interaction strengths, the regime
of primary physical interest is of course that of the strongly correlated 
Kondo lattice (KL): $n_f\rightarrow 1$, but with arbitrary conduction 
band filling $n_c$. The underlying low-energy model, obtained from the PAM
to leading order in $V^2$, is a Kondo lattice model (KLM); the KL regime 
of the PAM arising when $\ep_f=-|\ep_f|$ for $|\ep_f|/\Delta_0 \gg 1$
and $(U-|\ep_f|)/\Delta_0 \gg 1$, where $\Delta_0 \equiv \pi V^2 d^c_0(0)$
(with $d^c_0(0)$ the free ($V=0$) conduction band density of states at the 
Fermi level). The approach to the KL is not therefore unique, in that
$n_f\rightarrow 1$ arises for any asymmetry $\eta\equiv 1-2|\ep_f|/U
\in  [-1,1]$ on progressively increasing the interaction $U$. This is 
reflected in the fact that the associated KLM contains in general both
exchange {\em and} potential scattering contributions (the latter vanishes
as the asymmetry $\eta\rightarrow 0$ and is omitted in most studies of the KLM
{\it per se}, which thus correspond to symmetric $f$-levels alone but with
asymmetry retained in the conduction band).

Granted even a dominant interest is the strong coupling KL regime, the
resultant bare parameter space of the PAM (or KLM) nonetheless remains
`large', as above. The KL regime is however characterized by a low-energy
lattice scale, $\om_L$, diminishing progressively with increasing interaction
strength and expected to be exponentially small in strong coupling [7-25]. 
This scale is of course itself a function of the bare material parameters;
but that dependence is a subsidiary issue in comparison to the expectation
that physical properties of the PAM should exhibit universal scaling
in terms of $\om/\om_L$ and/or $T/\om_L$, in a manner largely 
independent of the bare parameters themselves. Understanding aspects
of such scaling behaviour will be a central theme of the present work.

 Our specific focus in this paper is on local single-particle dynamics
of the $T=0$ PAM,
embodied in 
$G^f_{ii}(\om)\;\leftrightarrow G^f_{ii}(t)= -\ii 
<\hat{T}(f^{\phantom{\dag}}_{i\sigma}(t) f^\dag_{i\sigma})>$ and 
likewise $G^c_{ii}(\om)$
for the $c$-levels, with corresponding local spectra $D^\nu_{ii}(\om)
=-\tfrac{1}{\pi}\sgn(\om)\Im\, G^\nu_{ii}(\om)$ (and $\nu=c$ or $f$). 

  We begin with some remarks on the trivial limit $V=0$ where
(Eq.(2.1))  the $f$-levels decouple from the free
conduction band. The latter is specified by its local propagator
denoted by 
 $g^c_0(\om)$, with corresponding density of states (dos)
$d^c_0(\om)\; (=N^{-1}\sum_\alpha\delta(\om - \ep_\alpha)$
with e.g.\ $\ep_\alpha\equiv \ep_\k$ for a Bloch decomposible
lattice); and it will 
prove useful to denote by $\H(z)$ the Hilbert transform
\beq
\H(z)=\int^\infty_{-\infty} d\ep \; \frac{\rho_0(\ep)}{z-\ep}
\label{eq:ht}
\eeq
for arbitrary complex $z$, where $\rho_0(\om)=d^c_0(\om; \ep_c=0)$ denotes
the free conduction band dos for $\ep_c=0$ (see Eq.(2.1)). 
The free $c$-electron propagator $g^c_0(\om)$
is then given by
\begin{subequations}
\beqa
g^c_0(\om)&=& \H(\om^+-\ep_c)  \\
&=&  \left[ \om^+ - \ep_c-S_0(\om) \right]^{-1}
\label{eq:s0def}
\eeqa
\end{subequations}
with $\om^+=\om + \ii\, \sgn(\om)0^+$ here and throughout, such that
$d^c_0(\om)=\rho_0(\om-\ep_c)$. Eq.(2.4b) simply
defines the Feenberg self-energy [39,40] used below,
with $S_0(\om)\equiv S[g^c_0]$ a functional of $g^c_0$ alone
(since $g^c_0=\H(S+1/g^c_0)$) from Eqs.(2.4)).
While our subsequent discussion holds for an essentially 
arbitrary $\rho_0(\om)$ and hence band structure embodied in $d^c_0(\om)$, 
specific results will later be given  
for the Bethe lattice (BL) and hypercubic lattice (HCL); for which
within DMFT the normalized $\rho_0(\ep)$ are respectively a semi-ellipse 
and an unbounded Gaussian, given explicitly ($t_*=1$) by [3-6] :
\begin{subequations}
\beqa
\rho_0(\ep)&=& \frac{2}{\pi } \left[ 1-\ep^2\right]^\frac{1}{2}
\;\;\;\;\;\; :\; |\ep| < 1 \;\;\;\;\;\,  \mbox{BL}  \label{eq:r0bl} \\
\rho_0(\ep)&=& \frac{1}{\surd{\pi} } \exp\left(-\ep^2\right)\;\;\,:
\;\; \;\hspace{1.3cm} \mbox{HCL } \label{eq:r0hc} 
\eeqa
\end{subequations}
Since $d^c_0(\om)=\rho_0(\om-\ep_c)$ is simply a rigid shift of
$\rho_0(\om)$ (the free conduction band is non-interacting), 
conduction band asymmetry is thus embodied in $\ep_c$ itself, with
$\ep_c=0$ the symmetric limit.

We turn now to the full local Green functions for the homogeneous
paramagnetic phase of interest, for which the $G^\nu_{ii}(\om)\equiv
G^\nu(\om)$ are site-independent. The essential simplifying feature of DMFT
-- and the key aspect of it as an approximation to finite-dimensional
systems -- is that the $f$-electron self-energy is site-diagonal 
(momentum independent) [3-6];
and from straightforward application of Feenberg renormalized perturbation 
theory [39,40]
the $G^\nu(\om)$ are given by
\begin{subequations}
\beqa
G^c(\om)&=& \left[ \om^+ - \ep_c-\frac{V^2}{\om^+ - \ep_f-\Sigma_f(\om)}
-S(\om) \right]^{-1} 
\label{eq:gc}  \\
G^f(\om)&=& \left[ \om^+ - \ep_f-\Sigma_f(\om)  - \frac{V^2}{ \om^+ 
-\ep_c-S(\om)} \right]^{-1} 
\label{eq:gf}
\eeqa
\end{subequations}
\beq
= \frac{1}{[\om^+-\ep_f-\Sigma_f(\om)]}\left\{ 1 + \frac{V^2}
{[\om^+-\ep_f-\Sigma_f(\om)]} G^c(\om) \right\}
\eeq
where $\Sigma_f(\om) = \Sigma_f^R(\om) - \ii\,\sgn(\om)\Sigma_f^I(\om)$ 
is the conventional single self-energy
(and the identity Eq.(2.7) follows from Eqs.(2.6)). 
The Feenberg self-energy 
$S(\om)\equiv S[G^c]$ is moreover precisely the same functional of 
the full $G^c(\om)$ as
it is of $g^c_0(\om)$ in the trivial limit of $V=0$ 
(e.g.\ $S=\tfrac{1}{4}t_*^2 G^c$ 
for the BL).  In consequence, $G^c(\om)$ is given directly using 
Eqs.(2.6a,4,3) by
\beq
G^c(\om)=\H(\gamma) \label{eq:gchg} 
\eeq
where
\beq
\gamma(\om)=\om^+ - \ep_c - \frac{V^2}{\om^+ - \ep_f - \Sigma_f(\om)}\,.
\label{eq:gamma}
\eeq

For an arbitrary conduction band $\rho_0(\ep)$, the approach to the full
interacting problem is clear in principle: given the self-energy
$\Sigma_f(\om)$, and hence $\gamma(\om)$ from Eq.(2.9), 
$G^c(\om)\\ =\H(\gamma)$ follows
directly from Hilbert transformation and $G^f(\om)$ from Eq.(2.7). 
But practice is another matter: the hard part is to find a suitable
approximation to the self-energy that can handle non-perturbatively 
the strongly correlated physics of the KL regime, as well as the weak coupling
regime of interactions (which itself is readily handled by plain perturbation
$\,$ theory $\,$ or $\,$ simple $\,$ variants thereof, see e.g.\ [14-17]). It is this impasse the
LMA seeks to break, via use of an underlying two-self-energy description
[26,27,34,35] as detailed in \S4. In addition 
of course the problem must be solved iteratively and self-consistently,
because an approximate $ \Sigma_f(\om)$ will itself be in 
general a functional of self-consistently determined propagators; 
that being a detail (albeit an important one) to which we likewise turn in 
\S4.3,4.

\subsection{Non-interacting limit}
\label{ssec:nil}

Before $\,$ proceeding $\,$ we $\,$ comment $\,$ briefly $\,$ on the non-interacting (NI) limit,
$U=0$, the local propagators for which are denoted by $G^\nu_0(\om)$
(or $G^\nu_0(\om;\, \ep_c, \ep_f, V^2)\;$ if explicit dependence on the
bare parameters is required) with corresponding spectra $D^\nu_0(\om)$.
Itself trivially soluble, the importance of the NI limit and rationale for 
discussing it, resides in its connection to the fully interacting problem 
via both Luttinger's theorem [36] and the quasiparticle behaviour of the full
$D^\nu(\om)$ at sufficiently low $\om$; as considered in \S3 below. The
$G^\nu_0(\om)$ are given by Eqs.(2.6-9) with $\Sigma_f=0$ and 
$\gamma(\om)\rightarrow \gamma_0(\om)=\om^+-\ep_c-V^2/[\om^+-\ep_f]$, with 
resultant spectra
\begin{subequations}
\beqa
 D^c_0(\om)&=&\rho_0(\gamma_0) = \rho_0\left(\om-\ep_c-\frac{V^2}
{\om-\ep_f}\right)\\
 D^f_0(\om)&=&\frac{V^2}{(\om-\ep_f)^{2}} \rho_0\left(\om-\ep_c-\frac{V^2}
{\om-\ep_f}\right)
\eeqa
\end{subequations}
and hence total band filling $n_{tot}=n^0_c+n^0_f$ given by
\beqa
\frac{1}{2}(n^0_c+n^0_f)& = \int^0_{-\infty} d\om\, [D^c_0(\om)+D^f_0(\om)]\nnu
\\
&=\int_{-\infty}^{-\ep_c+1/\eft} \rho_0(\ep)\,d\ep  + \theta(-\eft)
\eeqa
where $\eft = \ep_f/V^2\;$ (and $\theta(x)$ is the unit step function).
For band fillings $n_{tot}
\in (0,4)$ the system is generically metallic, with a non-zero Fermi level
dos $D^\nu_0(\om=0)$. But 
 since $\gamma_0$ diverges as $\om \rightarrow \ep_f$, 
the spectral functions in the vicinity of $\om=\ep_f$ 
have the same behaviour
 as the tails of the bare conduction band $\rho_0(\ep)$. So for a typical
bounded $\rho_0(\ep)$, e.g.\ the BL Eq.(2.5a), a hard spectral gap
 opens up in the neighbourhood
 of $\ep_f$; while for an unbounded $\rho_0(\ep)$, e.g.\ the HCL,  
a (strictly) soft gap arises at $\om=\ep_f$. The system is of course
 insulating -- a well known hybridization gap insulator [41] -- only if
the Fermi level ($\om=0$) lies in the gap (excluding the trivial case
of wholly empty or full bands); and from Eqs.(2.10,11)
the sufficient condition for this to occur is readily seen to be $n_{tot}
=2$, i.e.\ half-filling, which holds also for the fully interacting 
problem now considered.

\section{Adiabatic continuity, and quasiparticle behaviour.}

On increasing the interaction $U$ from zero the system remains
perturbatively connected to the NI limit; i.e.\ is a Fermi liquid,
a statement applicable both to the metallic heavy Fermion (HF) state
and the Kondo insulating (KI) phase which likewise evolves continuously
from the non-interacting hybridization gap insulator [34,35].
This adiabatic continuity requires that the Luttinger integral vanish [2,36],
{\it i.e.\ }that
\beq
I_L=\Im\int^0_{-\infty} \frac{d\om}{\pi}\, \frac{\partial \Sigma_f(\om)}
{\partial\om}\, G^f(\om)=0
\eeq

What may be deduced on entirely general grounds from $I_L=0\,$?
To that end note first that the local propagators $G^\nu(\om)\,(\nu=c\,$ or
$f$) may be expressed as 
\begin{subequations}
\beq
G^\nu(\om)=\int^\infty_{-\infty}d\ep\,\rho_0(\ep)\,G^\nu(\ep;\om)\,.
\eeq
The $c$-electron $G^c(\ep;\om) (\equiv N^{-1} \sum_\k
[\om^+ - \ep_c -\ep_\k - \Sigma_c(\om)]^{-1}$ with $\ep_\k\equiv \ep$) 
follows directly from Eqs.(2.8,3) as
\beq
G^c(\ep;\om)=\left[\om^+ - \ep_c - \frac{V^2}{\om^+ - \ep_f - \Sigma_f(\om)}
-\ep \right]^{-1}
\eeq
while $G^f(\ep;\om)$ follows in turn using Eq(2.7) as
\beq
G^f(\ep;\om) = \left[\om^+ -\ep_f - \Sigma_f(\om)- \frac{V^2}{\om^+ - 
\ep_c -\ep}\right]^{-1}\,.
\eeq
\end{subequations}
And in terms of the $G^\nu(\ep;\om)$ note that the total band filling
$n_{tot} = n_c+n_f$ is given generally by
\beq
\begin{split}
\tfrac{1}{2}(n_c+n_f)=&\Im\int^{\infty}_{-\infty} d\ep\,\rho_0(\ep)\,\times\\
&\int^0_{-\infty}\frac{d\om}{\pi} \left[ G^c(\ep;\om) + G^f(\ep;\om)\right]\,.
\end{split}
\eeq
Now use Eqs.(3.2a,c) in Eq.(3.1), $I_L=0$, together with the identity
 (from Eq.(3.2))
\beq
\begin{split}
\frac{\partial \Sigma_f(\om)}{\partial\om}\,& G^f(\ep;\om)=
\left[ G^c(\ep;\om) + G^f(\ep;\om)\right] - \\
& \frac{\partial}{\partial\om}\,
\ln\left[(\om^+ - \ep_c -\ep)(\om^+ -\ep_f - \Sigma_f(\om)) - V^2\right]
\end{split}
\nnu
\eeq
and perform the $\om$-integration. This yields 
\beq
\tfrac{1}{2}(n_c+n_f)=\int^{\infty}_{-\infty}\frac{d\ep}{\pi}\,
\rho_0(\ep)\,g(\ep)
\eeq
using only that $\Sigma_f^I(\om=0)=0$, holding for both the HF and KI
states; where \\ $g(\ep)=\tan^{-1}\left(\eta(\ep+\ep_c+\ep_f^*)/
[\ep_f^*(\ep+\ep_c) - V^2]\right)$ \\(with $\eta=0^+$), and the renormalized
level
\beq
\ep_f^*=\ep_f + \Sigma_f^R(\om=0)
\eeq
is thus defined. The $\ep$-integration in Eq.(3.4) is then readily
performed to give the desired result
\beq
\tfrac{1}{2}(n_c+n_f)=\int^{-\ep_c+1/\eft^*}_{-\infty}\rho_0(\ep)\,d\ep
\;\;+\;\;\theta(-\eft^*)
\eeq
where $\eft^*=\ep_f^*/V^2$.

Eq.(3.6) is equivalently a statement of Luttinger's theorem for the Fermi
surface of the PAM, for the relevant case of a local, momentum
independent self-energy appropriate to DMFT (the Fermi surface
is of course ``large'', including $f$- and $c$- electrons, see also [38]).
Three points should be noted about Eq.(3.6). (i) First and most
importantly we see it to be exact, following directly from $I_L=0$
without further approximation. (ii) It amounts to a simple renormalization
of the NI limit result Eq.(2.11); being of just that form but with
the bare level $\eft=\ep_f/V^2$ replaced by the renormalized level
$\eft^* = [\ep_f+\Sigma^R_f(0)]/V^2$; which is thus determined via 
Eq.(3.6) for given filling $n_{tot}$ (and $\ep_c$). (iii) Eq.(3.6)
is the direct analogue for the PAM of the Friedel sum rule
for an AIM [2,42], which relates the excess impurity charge 
($n_{imp}$) to the renormalized impurity level $\ep^*_{imp}$, and
which likewise follows directly from $I_L=0$ for the impurity model; see
also \S4.5.
In physical terms that parallel is entirely natural, given the connection to
an {\em effective} impurity model which is inherent to DMFT [3-6]
(Eq.(2.6b) for $G^f(\om)$ being of effective ``single-impurity''
form $G^f(\om) = [\om^+ - \ep_f -\Sigma_f(\om) - \Delta_{eff}(\om)]^{-1}$
with an effective, $\om$-dependent  hybridization $\Delta_{eff}(\om)
=V^2[\om^+-\ep_c - S(\om)]^{-1}$). Finally, we add that imposition of
Eq.(3.6) as a self-consistency condition will play an important
role in the LMA developed in \S4ff.

The second key implication of adiabatic continuity is that the limiting
low-$\om$ behaviour of the propagators $G^\nu(\om)$ amount to a 
renormalization of the NI limit, which is of course the origin of
the renormalized band picture [2,43]. This follows simply by
employing the leading low-$\om$ expansion of $\Sigma_f(\om)$,
\beq
\Sigma_f(\om)\sim \Sigma^R_f(0) - \left[\frac{1}{Z} - 1\right]\om
\eeq
with $Z=[1-(\partial\Sigma_f^R(\om)/\partial\om)_{\om=0}]^{-1}$
the quasiparticle weight/ mass renormalization and $\Sigma^I_f(\om)$
neglected as $\om\rightarrow 0$ (since $\Sigma^I_f(\om) \propto
\om^2$ for the HF metals or vanishes in the gap for the KI case). The 
low-$\om$ behaviour of the $G^\nu(\om)$ then follow from Eqs.(2.6)
as
\begin{subequations}
\beqa
G^c(\om)&\sim G^c_0(\om;\ep_c, Z\ep_f^*, ZV^2) \equiv \tilde{G}^c(\om) \\
G^f(\om)&\sim G^f_0(\om;\ep_c, Z\ep_f^*, ZV^2) \equiv Z\tilde{G}^f(\om) 
\eeqa
\end{subequations}
with $G_0^\nu$ the NI propagators (\S 2.1) and the quasiparticle
Green functions thus defined; with corresponding spectra
\begin{subequations}
\beq
D^c(\om)\sim \rho_0(\om - \ep_c - \frac{ZV^2}{(\om - Z\ep_f^*)}
) \equiv \tilde{D}^c(\om) 
\eeq
\beq
\begin{split}
D^f(\om)\sim \frac{Z^2V^2}{(\om - Z\ep_f^*)^2}\,\rho_0
(\om - &\ep_c - \frac{ZV^2}{(\om - Z\ep_f^*)})\\
&\equiv Z\tilde{D}^f(\om) 
\end{split}
\eeq
\end{subequations}
($\ep_f^*$ is the renormalized level, Eq.(3.5)). And the total band
filling $n_{tot}=n_c+n_f$, calculated from the quasiparticle 
propagators as $\tfrac{1}{2}n_{tot} = \int^0_{-\infty}d\om\,
[\tilde{D}^c(\om) + \tilde{D}^f(\om)]$, correctly satisfies the 
exact result Eq.(3.6).

Eqs(3.8,9) embody the quasiparticle behaviour of the PAM, and 
have important implications for the scaling
behaviour of $D^\nu(\om)$ in the strong coupling regime of primary 
interest, as now considered. In the KL regime where $n_f\rightarrow 1$,
the quasiparticle weight $Z$ becomes exponentially small (as considered
explicitly in \S5.2). Defining a low-energy lattice coherence scale
by ($t_*=1$)
\beq
\om_L=ZV^2
\eeq
the scaling behaviour of dynamics corresponds to considering
finite $\om'=\om/\om_L$ in the formal limit $\om_L\rightarrow 0$.
Eqs.(3.9) then yield 
\begin{subequations}
\beqa
D^c(\om)&\sim& \rho_0(- \ep_c - \frac{1}{(\om' - \eft^*)})\\
V^2 D^f(\om)&\sim& \frac{1}{(\om' - \eft^*)^2}\,\rho_0
(- \ep_c - \frac{1}{(\om' - \eft^*)})
\eeqa
\end{subequations}
where `bare' factors of $\om\equiv \om_L\om'$ may be neglected, and 
$\eft^*=\ep_f^*/V^2$. Moreover, in the KL regime, $\eft^*$ is solely dependent 
upon $\ep_c$: from Eq.(3.11b) with $\tfrac{1}{2}n_{f} = \int^0_{-\infty}
d\om \tilde{D}^f(\om)$ $(\tilde{D}^f(\om)=D^f(\om)/Z)$,
\beqa
\tfrac{1}{2}n_{f} &=& \int^0_{-\infty}d\om' \,\frac{1}{(\om' - \eft^*)^2}\,\rho_0
(- \ep_c - \frac{1}{(\om' - \eft^*)}) \nnu \\
&=&\int^{-\ep_c+1/\eft^*}_{-\ep_c} d\ep\,\rho_0(\ep)\;\;+\;\;\theta(-\eft^*)
\eeqa
whence $\eft^*\equiv\eft^*(\ep_c)$ as $n_f\rightarrow 1$ (and in addition
$\sgn(\eft^*) = \sgn(\ep_c)$).

Eqs.(3.11) embody the low-$\om$ behaviour of the single-particle spectra
$D^\nu(\om)$, in the KL regime where $n_f\rightarrow 1$ but for 
arbitrary conduction band filing $n_c$; regarding which the following
important points should be noted. (i) Eqs.(3.11) show that both
$D^c(\om)$ and $V^2D^f(\om)$ (and not therefore $D^f(\om)$ itself)
exhibit one-parameter universal scaling in terms of $\om'=\om/\om_L$:
with {\em no} explicit dependence on the bare material parameters
$U, \eta$ (or $\ep_f$) and $V^2$; and dependent solely upon $\ep_c$
(or equivalently on the conduction band filling $n_c\equiv n_c(\ep_c)$,
see below) which itself determines the renormalized level 
$\eft^*\equiv\eft^*(\ep_c)$ as above. (ii) Eqs.(3.11) provide 
explicitly the limiting behaviour that, as $|\om'|=|\om|/\om_L\rightarrow 0$,
must of necessity be recovered by any credible microscopic theory; and direct comparison of LMA results to which will be given in \S5.3. Of equal 
importance however, the simple results above are asymptotically valid
only as $\om'\rightarrow 0$, and prescribe neither the $\om'$-range over which
Eqs.(3.11) hold nor the general $\om'$-dependence of the scaling spectra --
for which a real theory  is required. (iii) The particle-hole symmetric PAM discussed in [34] (for which $n_f=1=n_c$) is just a particular case of
the above, in which $\ep_c=0$ and the renormalized level $\eft^*=0$
(by symmetry); and where the low-energy lattice scale $\om_L$
(Eq.(3.10)) is precisely the gap scale characteristic of the Kondo
insulating state [34,35]. 
Finally, scaling arguments {\it per se}
are obviously independent of how the low-energy KL scale $\om_L
\equiv \om_L(\ep_c, U, \eta, V^2)$ itself depends upon the bare parameters;
an issue of intrinsic interest that has long attracted attention
(see e.g.\ [8,21-25]) but which we believe (as argued in \S5) to be in large 
part a red herring in understanding the expected connection
between the KL regime of the PAM, and single-impurity Kondo physics, on 
suitably large energy and/or temperature scales.

Before proceeding to the LMA we mention one further 
implication of Eq.(3.12) applicable to the KL regime: together with the exact
result Eq.(3.6) it gives 
\beq
\tfrac{1}{2}n_c=\int^{-\ep_c}_{-\infty}d\ep\,\rho_0(\ep)
\eeq
for the $c$-band filling. This shows (a) that $n_c\equiv n_c(\ep_c)$ is
indeed determined by $\ep_c$ as noted above; and (b) that the resultant 
$n_c$ is just that for the free ($V=0$) conduction band, for which (\S 2)
$d^c_0(\om)=\rho_0(\om-\ep_c)$ with $\tfrac{1}{2}n_c=\int^{0}_{-\infty}
d\om\,d^c_0(\om)$. In physical terms this is natural, since from 
Eq.(3.8a) the effective hybridization is $ZV^2=\om_L$, exponentially
small in the KL regime such that the net conduction band filling is in effect
independent of coupling to the $f$-levels. We shall comment further on the
latter in \S5.

\section{Local Moment Approach}
  The discussion thus far has been couched in terms of the
single self-energy $\Sigma_{f}(\om)$ which, via diagrammatic 
perturbation theory in the interaction strength, provides 
the conventional route to dynamics. A determination of the 
propagators in this way is not however mandatory.  Indeed while 
fine in principle there are good reasons to avoid it; stemming
from the practical inability of conventional perturbation theory, or partial
resummations thereof, to handle the strongly correlated regime of primary
interest. For this reason the LMA [26-35] takes a different route to 
the problem, employing instead a two-self-energy description that 
is a natural consequence of                                        
the mean-field description from which it starts. Here we first consider the
implications of such in general terms, independent of
subsequent details of implementation (\S4.3,4 ) and not confined to the
symmetric PAM considered hitherto [34,35].

  There are three essential elements to the LMA [26-35]. (i) Local moments
(`$\mu$'), regarded as the first effect of interactions, are introduced 
explicitly and self-consistently from the outset. The starting point is 
thus simple broken symmetry static
mean-field (MF, {\it i.e.\  }unrestricted Hartree-Fock); containing two degenerate,
locally symmetry broken MF states corresponding to $\mu = \pm |\mu|$ [34].
 While severely deficient by itself (see {\it e.g.\ }[26,27,31,34] and below), 
MF nonetheless provides a suitable starting point for a non-perturbative 
many-body approach to the problem.
(ii) To this end the LMA employs a two-self-energy description that follows
naturally from the underlying two local MF saddle points; with associated 
dynamical
self-energies built diagrammatically from, and functionals of, the 
underlying MF propagators.
(iii) The third and most important idea behind the LMA is that of symmetry
restoration [27,28,32,34]: self-consistent restoration of the broken symmetry inherent 
at pure MF level; and in consequence, as discussed below, correct recovery 
of the local Fermi liquid behaviour that reflects adiabatic continuity in 
$U$ to the non-interacting limit.

Within a two-self-energy description
the local propagators $G^{\nu}(\om)$, which are correctly rotationally 
invariant, are expressed formally as
({\it cf}  Eqs.(2.6))
\beq
G^{\nu}(\om) = \tfrac{1}{2}[G^{\nu}_{\upa}(\om) + G^{\nu}_{\dna}(\om)]
\eeq
where (with $\sigma = \upa/\dna$ or $+/-$)
\begin{subequations}
\beq
G^{c}_{\sigma}(\om) = [\om^{+}-\ep_{c} - \frac{V^{2}}
{\om^{+}-\ep_{f} - \tilde{\Sigma}_{\sigma}(\om)} -S(\om)]^{-1} 
\eeq
\beq
G^{f}_{\sigma}(\om) = [\om^{+} -\epsilon_{f} - \tilde{\Sigma}_{\sigma}(\om)
-\frac{V^{2}}{\om^{+}-\ep_{c}-S(\om)}]^{-1}
\eeq
\end{subequations}
and $S(\om) \equiv S[G^{c}]$ as usual (the reader is referred to [34] for further,
physically oriented discussion of these basic equations).
The $f$-electron self-energies $\tilde{\Sigma}_{\sigma}(\om)$ are conveniently
separated as
\beq
\tilde{\Sigma}_{\sigma}(\om) = \tfrac{U}{2}(\bar{n} - \sigma |\bar{\mu}|)
+ \Sigma_{\sigma}(\om)
\eeq
where the first term represents the purely static Fock bubble diagram 
which alone is retained at pure MF level (with $\bar{n}$ and $|\bar{\mu}|$ 
given explicitly by 
Eq.(4.12) below); and where the second term, $\Sigma_{\sigma}(\om) =
\Sigma^{R}_{\sigma}(\om) - \ii\, \sgn(\om)\Sigma^{I}_{\sigma}(\om)$, is the key
dynamical contribution mentioned above (`everything post-MF').

  Eqs.(4.1,2) are the direct counterparts of the single self-energy equations
Eqs. (2.6) (to which they would trivially reduce if $\tilde{\Sigma}_{\sigma}(\om) 
\equiv \Sigma_{f}(\om)$ for each $\sigma$). For an arbitrary conduction
band dos $\rho_{0}(\om)$ and any given $\{\tilde{\Sigma}_{\sigma}(\om)\}$,
they are likewise readily solved ({\it cf } the discussion of 
Eqs.(2.6a,\\ 9)): defining
\beq
\gamma_{\sigma}(\om) = \om^{+} - \ep_{c} - \frac{V^{2}}{\om^{+}
-\ep_{f} - \tilde{\Sigma}_{\sigma}(\om)}
\eeq
such that $G^{c}(\om) =\tfrac{1}{2}\sum_{\sigma}[\gamma_{\sigma}-S]^{-1}$
(Eqs.(4.1,2a)), and comparing to $G^{c}(\om) = [\gamma -S]^{-1}$ (Eqs.(2.6a,9)),
the $\gamma_{\sigma}$'s are related to the single $\gamma(\omega)$ (Eq.(2.9)) by
\beq
\gamma(\om) = \tfrac{1}{2}[\gamma_{\upa}(\om) + \gamma_{\dna}(\om)]
+ \frac{[\tfrac{1}{2}(\gamma_{\upa}(\om) -\gamma_{\dna}(\om))]^{2}}
{S(\om) - \tfrac{1}{2}[\gamma_{\upa}(\om) + \gamma_{\dna}(\om)]}.
\eeq
Given $\tilde{\Sigma}_{\sigma}(\om)$ and hence $\gamma_{\sigma}(\om)$, 
this equation together with
\beq
S(\om) = \gamma - \frac{1}{H(\gamma)}
\eeq
(from $G^{c}(\om) =[\gamma -S]^{-1}$ and Eq.(2.8)) may be solved 
straightforwardly and rapidly in an iterative fashion; employing 
an initial `startup' for $S$ (typically
$S = \tfrac{1}{4}g_{0}(\omega)$ with $g_{0}$ the free conduction band propagator
with dos $\rho_{0}(\om)$). With $S(\om)$ then known, the $G^{\nu}(\om)$
follow directly from Eqs.(4.1,2).

  The conventional single self-energy $\Sigma_{f}(\om)$ follows immediately,
essentially as a byproduct, because solution of Eqs.(4.5,6) determines 
both $S(\om)$ {\em and}
$\gamma(\om)$, whence (from Eq.(2.9)) $\Sigma_{f}(\om) = \om^{+} - \ep_{f}
-V^{2}[\om^{+}- \ep_{c} -\gamma(\om)]^{-1}$ follows; which relation
may be expressed equivalently as
\beq
\Sigma_{f}(\om) = \tfrac{1}{2}[\tilde{\Sigma}_{\upa}(\om)+
\tilde{\Sigma}_{\dna}(\om)]+
\frac{[\tfrac{1}{2}(\tilde{\Sigma}_{\upa}(\om)-\tilde{\Sigma}_{\dna}
(\om))]^{2}}
{{\cal{G}}^{-1}(\om)-\tfrac{1}{2}[\tilde{\Sigma}_{\upa}(\om) +
\tilde{\Sigma}_{\dna}(\om)]}.
\eeq
Here ${\cal{G}}(\om)$ is the usual host/medium propagator [44], given by
${\cal{G}}^{-1}(\om) =
[(G^{f}(\om))^{-1} + \Sigma_{f}(\om)] = [\om^{+} -\ep_{f}
-V^{2}(\om^{+} -\ep_{c} - S(\om))^{-1}]$ with corresponding
spectral density ${\cal{D}}(\om) = -\tfrac{1}{\pi}\sgn(\om)\Im{\cal{G}}
(\om)$ (and which in physical terms includes interactions on {\em all } sites
other than the local site
$i$ (${\cal{G}}(\om) \equiv {\cal{G}}_{ii}(\om)$) [44]).
 The conventional single self-energy may thus be obtained directly
given the $\{\tilde{\Sigma}_{\sigma}(\om)\}$, although obviously not {\it vice
versa}, and the underlying two-self-energy description may be viewed equivalently
as a means to obtain $\Sigma_{f}(\om)$. The particular class of diagrams
retained in practice for the dynamical $\{\Sigma_{\sigma}(\om)\}$ 
(see Eq.(4.3)) will be detailed in \S 4.3; at present none need be specified.

  At the pure MF level of unrestricted Hartee-Fock, dynamical contributions
to the $\tilde{\Sigma}_{\sigma}(\om)$ are of course neglected entirely and
$\tilde{\Sigma}_{\sigma}(\om) \equiv \tilde{\Sigma}^{0}_{\sigma} =
\tfrac{U}{2}(n-\sigma|\mu|)$ (with the MF local $f$-level charge $n \equiv
\bar{n}$ and moment $|\mu| \equiv |\bar{\mu}|$ determined in the 
usual simple fashion, \S 4.2).  From Eq.(4.7) the single self-energy at 
MF level is then
\beq
\Sigma_{f}^{MF}(\om) = \tfrac{1}{2}Un + \frac{(\tfrac{1}{2}U|\mu|)^{2}}
{{\cal{G}}_{0}^{-1}(\om) - \tfrac{1}{2}Un}
\eeq
(with ${\cal{G}}_{0}(\om)$ the corresponding MF medium propagator, whose
Fermi level spectral density ${\cal{D}}_{0}(\om =0)$ is readily shown to
be non-zero).  From this the basic deficiency of pure MF is clear:
if the local moment $|\mu| \neq 0$, then from Eq.(4.8) the Fermi level
$\Im\Sigma_{f}^{MF}(\om =0) \neq 0$ and Fermi liquid behaviour is violated
--- wholly wrong, albeit arising naturally because the resultant degenerate
MF local moment state is not perturbatively connected to the non-interacting
limit. While this problem would not occur if $|\mu| =0$ were enforced 
{\em a priori } (restricted
Hartree-Fock), another one then arises at post-MF level; for from Eq.(4.8) the
two- and single- self-energy descriptions then coincide, with 
$\Sigma_{f}^{MF}(\om) = \tfrac{1}{2}Un$ merely the static Hartree 
contribution, producing a trivial energy shift to the non-interacting 
propagators.  Subsequent construction of the dynamical $\Sigma_{f}(\om)$ 
via conventional perturbation theory in $U$ employing these propagators,
is equivalent to expanding about the restricted Hartree-Fock saddle-point. 
But when local moments can form at MF
level this single-determinantal saddle point, unlike its unrestricted 
MF counterpart, is {\em unstable } to particle-hole excitations. It is this 
in turn that is readily shown to underlie
the familiar divergences arising within conventional perturbation theory if
one attempts to perform the `natural' diagrammatic resummations (such as RPA)
that one expects physically are required to capture regimes of strong 
electronic correlation, and the general inability to surmount which has 
been a plague on all our houses [2,45].

  The LMA seeks to surmount these problems by (a) retaining the two-self-energy
description, with the inherent notion of local moments and essential stability
of the underlying MF state; while (b) incorporating many-body dynamics into
the associated self-energies $\{\tilde{\Sigma}_{\sigma}(\om)\}$ in a simple
and tractable fashion, and in such a way that Fermi liquid behaviour is 
recovered at low-energies.

\subsection{ Symmetry restoration }

  This brings us to the key notion of symmetry restoration (SR), now sketched
briefly in the generic context of heavy Fermion (HF) metals in the asymmetric 
PAM, where it arises from the obvious question: under what conditions on the
$\{\tilde{\Sigma}_{\sigma}(\om)\}$ will the $f$-electron single self-energy
$\Sigma_{f}(\om)$ exhibit Fermi liquid behaviour as $\om \rightarrow 0$,
{\it i.e.\ }will $\Sigma_{f}^{I}(\om) \sim {\cal{O}}(\om^{2})$? This may be 
answered simply by employing a general 
low-frequency Taylor expansion for the $\tilde{\Sigma}_{\sigma}(\om)$
in Eq.(4.7), along precisely the same lines as in [27] for the Anderson impurity
model. That is merely a matter of algebra, and from it one finds the
necessary/sufficient condition for $\Sigma_{f}^{I}(\om) \sim {\cal{O}}
(\om^{2})$ is that
\beq
\tilde{\Sigma}^{R}_{\upa}(\om =0) = \tilde{\Sigma}^{R}_{\dna}(\om =0).
\eeq
Moreover, with Eq.(4.9) satisfied then from Eq.(4.7) (i) all self-energies
coincide at the Fermi level, {\it i.e.\ } 
\beq
\Sigma^{R}_{f}(\om =0) = \tilde{\Sigma}^{R}_{\sigma}(\om =0)
\eeq
for either spin $\sigma$; (ii) the leading low-$\om$ behaviour of
$\Sigma_{f}^{R}(\om)$ is as in Eq.(3.7), with the quasiparticle weight
$Z=[1-(\partial \Sigma^{R}_{f}(\om)/\partial\om)_{\om =0)}]^{-1}$
given by $Z^{-1} = \tfrac{1}{2}(Z^{-1}_{\upa}+Z^{-1}_{\dna})$
where $Z_{\sigma} =
[1-(\partial{\Sigma}^{R}_{\sigma}(\om)/\partial\om)_{\om =0}]^{-1}$
is thus defined; and (iii) the quasiparticle behaviour embodied in
Eqs.(3.8) for the propagators $G^{\nu}(\om)$ is thus recovered.

  Eq.(4.9), a condition upon the $\tilde{\Sigma}_{\sigma}^{R}(\om =0)$ solely
{\em at } the Fermi level, is the SR condition that is central to the LMA.
It is quite general, being precisely the condition found for Anderson 
impurity models,
whether metallic [26,27] or pseudogap AIMs [31,32]; and likewise 
for the particle-hole symmetric limit of the PAM [34] where it guarantees 
persistence of the insulating gap with increasing interaction strength, 
reflecting the `insulating Fermi liquid' nature of the Kondo insulating state.
The general
consequences of SR are correspondingly common to all these problems:
In practice, Eq.(4.9) amounts to a self-consistency equation for the
local moment $|\mu|$ (supplanting the pure MF condition for $|\mu|$),
see \S4.3,4.  Most importantly, \S 4.4 and [26-28,34], imposition of SR as a 
self-consistency condition generates a non-vanishing low-energy 
spin-flip scale $\om_{m}$, manifest in 
particular in the transverse spin polarization propagator, whose physical 
significance is that it sets a non-vanishing timescale, $\tau \sim h/\om_{m}$,
for the restoration of the broken spin-symmetry endemic to the pure MF level of
description; and which in the present context is equivalently the low-energy
Kondo lattice scale, with $\om_{m} \propto \om_{L} = ZV^{2}$ (Eq.(3.10)).

\subsection{ Mean-field }

Since the self-energies $\tilde{\Sigma}_{\sigma}(\om)$ are built
diagrammatically from the underlying MF propagators, it is appropriate at this 
stage to comment briefly on MF itself; denoting the MF propagators
by $g^{\nu}_{\sigma}(\om)$ (and $g^{\nu}(\om) =
\tfrac{1}{2}\sum_{\sigma}g^{\nu}_{\sigma}(\om)$). These follow
from Eqs.(4.2) as
\begin{subequations}
\beqa
g^{c}_{\sigma}(\om) &=& [\om^{+} - \ep_{c} - \frac{V^{2}}
{\om^{+}-e_{f} +\sigma x} -S(\om)]^{-1}  \\
g^{f}_{\sigma}(\om) &=& [\om^{+} -e_{f}+\sigma x - \frac{V^{2}}
{\om^+ - \epsilon_{c} -S(\om)}]^{-1}
\eeqa
\end{subequations}
where $S(\om) \equiv S[g^{c}]$; and we have written the purely static
$\ep_{f}+\tilde{\Sigma}_{\sigma}(\om) \equiv \ep_{f} +
\tilde{\Sigma}^{0}_{\sigma}$ as
$\ep_{f} + \tilde{\Sigma}^{0}_{\sigma} =
e_{f} - \sigma x$, with $x = \tfrac{1}{2}U|\mu|$ and $e_{f}$ given at pure
MF level by $e_{f} = \epsilon_{f} + \tfrac{1}{2}Un$.
For any given $e_{f}$ and $x$, explicit solution of Eqs.(4.11) for the MF 
propagators $g^{\nu}_{\sigma}(\om) \equiv g^{\nu}_{\sigma}(\om; e_{f},x)$
follows directly in one shot as described above (Eqs.(4.4-6)). And at pure 
MF level, the local moment $|\mu|$ and charge $n$ are found from the 
usual MF self-consistency conditions $|\mu| =|\bar{\mu}|$ and $n=\bar{n}$; 
where
$|\bar{\mu}| \equiv |\bar{\mu}(e_{f},x)|$ and $\bar{n}\equiv \bar{n}(e_{f},x)$
are given generally by
\begin{subequations}
\beqa
|\bar{\mu}| &=& \int^{0}_{-\infty} ~d\om ~~[d^{f}_{\upa}(\omega;e_{f},x)-
d^{f}_{\dna}(\om;e_{f},x)]\\
\bar{n} &=& \int^{0}_{-\infty} ~d\om ~~[d^{f}_{\upa}(\om;e_{f},x)+
d^{f}_{\dna}(\om;e_{f},x)]
\eeqa
\end{subequations}
(such that the static Fock bubble diagram, appearing in Eq.(4.3) for
$\tilde{\Sigma}_{\sigma}(\om)$, is given generally by $\tfrac{U}{2}(\bar{n}
-\sigma|\bar{\mu}|)$). In practical terms here it is obviously
most efficient to work with fixed $e_{f}$ and $x$: from Eq.(4.12a),
$|\mu| = |\bar{\mu}|$ yields immediately $U=2x/|\mu|$, whence
with $n = \bar{n}$ Eq.(4.12b) likewise gives directly 
$\ep_{f} =e_{f} -\tfrac{U}{2}n$. One can of course choose equivalently 
to specify the bare parameters $U$ and $\ep_{f}$ (or
$\eta = 1+2\ep_{f}/U$) from the beginning --- which simply requires
iterative cycling of the pure MF self-consistency equations --- whence
Eqs.(4.12) determine the pure MF values for $x=\tfrac{1}{2}U|\mu|$ and $e_{f}$.
Results arising at pure MF level will be shown explicitly in \S5.1.

\subsection{ LMA: practice }

  Beyond the crude level of pure MF it is of course the dynamical contributions
to the self-energies, $\Sigma_{\sigma}(\om)$ (Eq.(4.3)), that are all 
important; and since the $\Sigma_{\sigma}(\om)$ are functionals of the 
underlying MF $f$-electron propagators $\{g^{f}_{\sigma}(\om;e_{f},x)\}$ 
themselves given by Eq.(4.11b), the 
$\tilde{\Sigma}_{\sigma}(\om) \equiv \tilde{\Sigma}_{\sigma}(\om; e_{f},x)$
thus depend upon $e_{f}$ and $x$. Hence, independently of the
particular class of diagrams retained in practice for the dynamical
self-energies, the question arises as to how $x=\tfrac{1}{2}U|\mu|$ and 
$e_{f}$ are determined in general (for any given set of bare model parameters)?
To do so clearly requires two conditions. As discussed in \S 4.1, the symmetry
restoration condition Eq.(4.9) must of necessity be satisfied in a Fermi liquid
phase; and using Eq.(4.3) it may be cast as
\beq
\Sigma^{R}_{\upa}(\om =0; e_{f},x) - \Sigma^{R}_{\dna}(\om =0;e_{f},x)
= U |\bar{\mu}(e_{f},x)|.
\eeq
Likewise, as discussed in \S 3, adiabatic continuity requires that the 
Luttinger integral theorem Eq.(3.1) be satisfied, {\it i.e.\ }
\beq
I_{L}(e_{f},x) = \Im \int^{0}_{-\infty} ~ \frac{d\om}{\pi}
~ \frac{\partial\Sigma_{f}(\om)}{\partial\om} ~G^{f}(\om) = 0
\eeq
where the Luttinger integral itself depends necessarily on $e_{f}$ and $x$.
The essential point is obvious: these two equations are sufficient to determine
$x=\tfrac{1}{2}U|\mu|$ and $e_{f}$ in the general case, and in effect supplant
the corresponding pure MF self-consistency conditions discussed above 
(recall that since the broken symmetry MF state itself is not adiabatically 
connected to the non-interacting limit, MF fails in general to satisfy 
either symmetry restoration 
or the Luttinger theorem). The optimal method for solving these equations
will naturally  depend on the particular approximation employed for 
the dynamical $\{\Sigma_{\sigma}(\om)\}$; but that is an algorithmic detail, 
to which we return below. One further comment should be added here. 
It is readily shown that the particle-hole symmetric PAM considered in 
[34,35] (for which $\ep_{f} =-\tfrac{U}{2}, \ep_{c}=0$, see \S 2),
corresponds necessarily to $e_{f} =0$, and that the Luttinger
theorem Eq.(4.14) is automatically satisfied by particle-hole symmetry. In that
case solely the symmetry restoration condition is therefore required 
to determine $x=\tfrac{1}{2}U|\mu|$ and hence the local moment, $|\mu|$; 
precisely as employed in previous LMA work on Kondo insulators, [34,35]. 

  While the preceding discussion is general, the final task is  
to specify the class of diagrams contributing to the dynamical $f$-electron 
self-energies $\{\Sigma_{\sigma}(\om)\}$ that we here retain in practice.
These have the same functional form employed in [34,35] for the symmetric PAM,
and may be cast as shown in Fig.1. The wavy line denotes the local
interaction $U$, the double line denotes the broken symmetry host/medium 
propagator $\tilde{{\cal{G}}}_{-\sigma}(\om)$ specified below (Eq.(4.16)); and 
the local $f$-level transverse spin polarization propagator 
$\Pi^{-\sigma\sigma}(\om)$, likewise specified below, is shown as hatched. 
The diagrams translate to
\beq
\Sigma_{\sigma}(\om) = U^{2}\int^{\infty}_{-\infty} \frac{d\om_{1}}{2\pi \ii}
~ \tilde{{\cal{G}}}_{-\sigma}(\om -\om_{1}) \Pi^{-\sigma\sigma}(\om_{1})
\eeq
and retention of them is motivated on physical grounds, for they
describe correlated spin-flip processes that are essential in particular to
capture the strong coupling Kondo lattice regime: in which having, say, added
a $\sigma$-spin electron to a $-\sigma$-spin occupied $f$-level on lattice
site $i$, the $-\sigma$-spin hops off the $f$-level generating an on-site
spin-flip (with dynamics reflected in the polarization propagator
$\Pi^{-\sigma\sigma}(\om)$). The $-\sigma$-spin electron then propagates
through the lattice in a correlated fashion, interacting fully with
$f$-electrons on {\em any } site $j \neq i$ (as embodied in the host/medium
$\tilde{\cal{G}}_{-\sigma}(\om)$); before returning at a later time to the 
original site $i$, whence the originally added $\sigma$-spin is removed
(and which process simultaneously restores the spin-flip on site $i$).

  \hspace{0.5cm}The renormalized $f$-electron medium propagator
$\tilde{{\cal{G}}}_{-\sigma}(\om)$, which embodies correlated propagation of 
an $f$-electron through the lattice, is given explicitly by ({\it cf } its 
counterpart ${\cal{G}}(\om)$ arising in Eq.(4.7))
\beq
\tilde{\cal{G}}_{-\sigma}(\om) = [\omega^{+} - e_{f} -
\sigma x - \frac{V^{2}}{\om^{+}-\ep_{c} - S(\om)}]^{-1}
\eeq
with corresponding spectral density $\tilde{\cal{D}}_{-\sigma}(\om)$.
Physically, $\tilde{\cal{G}}_{-\sigma}(\om) \equiv 
\tilde{\cal{G}}_{ii;-\sigma}(\om)$
includes interactions on all sites other than $i$ (on which interactions occur
at MF level); and the dependence of $\Sigma_{\sigma}(\om)$ (Fig.1) on which
accounts in effect for the hard-core boson nature of the spin-flips [34,46]. 
Diagrammatic expansion of $\tilde{\cal{G}}_{-\sigma}(\om)$ in terms of MF
propagators and self-energy insertions $\Sigma_{-\sigma}(\om)$, and hence the
infinite set of diagrams implicit in Fig.1 for $\Sigma_{\sigma}(\om)$, is
discussed further in [34,46] to which the reader is referred.
\begin{figure}[t]
\epsfxsize=200pt
\centering
{\mbox{\epsffile{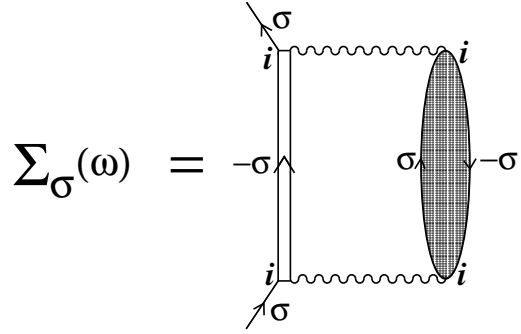}}}
\caption{
Diagrams retained in practice for the dynamical
$f$-electron self-energies $\Sigma_{\sigma}(\om)$. Wavy line:
interaction $U$. Double line: renormalized host/medium propagator
(see text). The transverse spin polarization propagator is shown 
hatched.
}
\end{figure}

  The local (site-diagonal) polarization propagator entering Eq.(4.15) for 
$\Sigma_{\sigma}(\om)$ is given at its simplest level by an RPA-like
particle-hole ladder sum in the transverse spin channel, namely
\beq
\Pi^{-\sigma\sigma}(\om) = {^{0}\!\Pi}^{-\sigma\sigma}(\om)
[1-U\, {^0\!\Pi}^{-\sigma\sigma}(\om)]^{-1}
\eeq
with the corresponding bare polarization bubble ${^{0}\!\Pi}^{-\sigma\sigma}
(\om)\\ \equiv  {^{0}\!\Pi}^{-\sigma\sigma}(\om; e_{f}, x)$ expressed in terms 
of the broken symmetry MF propagators $\{g^{f}_{\sigma}(\om;e_{f},x) \}$; 
referred to in [34] as `LMAI'.  [Alternatively, one may readily renormalize 
the polarization bubbles in terms of the host/medium propagators 
$\{ \tilde{\cal{G}}_{\sigma}(\om) \}$, so-called `LMAII' [34]. Results arising
from the two are however very similar [34], so we largely confine our attention
in the present paper to LMA I, excepting the explicit comparison between the
two made in Fig.9 below.] The ${^{0}\!\Pi}^{-\sigma\sigma}(\om)$ and
hence $\Pi^{-\sigma\sigma}(\om)$) are moreover $\,$ readily $\,$ shown $\,$ to
be related by $\;\Pi^{-\sigma\sigma}(\om) =\;\\ \Pi^{\sigma-\sigma}(-\om)$ [27,34];
whence only one such need be considered explicitly, say $\Pi^{+-}(\om)$.
Using this, and the Hilbert transform for $\Pi^{+-}(\om)$, Eq.(4.15) for
the dynamical self-energy reduces to 
\beqa
\Sigma_{\sigma}(\om) = U^{2}\int^{\infty}_{-\infty} \frac{d\om_{1}}{\pi} ~
&&\Im\Pi^{+-}(\om_{1})
\Bigl[\theta(\sigma\om_{1})\tilde{\cal{G}}_{-\sigma}^{-}(\om 
+ \sigma\om_{1}) \nnu \\ 
&&+ \theta(-\sigma\om_{1})\tilde{\cal{G}}_{-\sigma}^{+}(\om +
\sigma\om_{1})\Bigr]
\eeqa
where $\tilde{\cal{G}}_{-\sigma}^{\pm}(\om) =
\int^{\infty}_{-\infty}d\om_{1}\tilde{\cal{D}}_{-\sigma}(\om_{1})\theta
(\pm\om_{1})[\om-\om_{1} \pm \ii 0^{+}]^{-1}$ denote the one-sided Hilbert
transforms such that $\tilde{\cal{G}}_{-\sigma}(\om) =
\tilde{\cal{G}}^{+}_{-\sigma}(\om)
+\tilde{\cal{G}}^{-}_{-\sigma}(\om)$.

\subsection{Solution }

  We now summarise the preceding discussion from the viewpoint of practical
solution, and specify what we find to be a numerically efficient algorithm
to solve the basic LMA-DMFT equations.

  The self-energies $\{ \tilde{\Sigma}_{\sigma}(\om) \}$ are given in their
entirety by Eq.(4.3), with the static Fock contributions $|\bar{\mu}(e_{f},x)|$
and $\bar{n}(e_{f},x)$ from Eq.(4.12). The dynamical contribution to the
self-energy, $\Sigma_{\sigma}(\om)$, is given by Eq.(4.18), with the
polarization propagator therein specified by Eq.(4.17); and the host/medium 
propagator $\tilde{\cal{G}}_{-\sigma}(\om)$ given by Eq.(4.16) 
(itself dependent on the Feenberg self-energy $S(\om) \equiv S[G^{c}]$, 
requiring as such an iterative,
self-consistent solution of the problem). For given 
$\tilde{\Sigma}_{\sigma}(\om)$, Eqs.(4.4-6) are the key equations to solve
(as there discussed) for $G^{c}(\om)$ and $S(\om)$; $G^{f}(\om)$ then follows
directly from Eqs.(4.1,2b). In addition, centrally, both the symmetry 
restoration condition for $\tilde{\Sigma}_{\sigma}(\om =0)$ (Eq.(4.10) or
(4.13)) and the Luttinger integral theorem Eq.(4.14) --- or equivalently 
Eq.(3.6) --- must also be satisfied; which conditions, for given bare 
parameters $\{\ep_{c},V^{2},\ep_{f},U\}$, determine both 
$x = \tfrac{1}{2}U|\mu|$ (and hence the local moment $|\mu|$) and $e_{f}$ that 
prescribe the underlying MF propagators. In this regard we note for use
below that Eq.(4.4) for $\gamma_{\sigma}(\om)$ may be written equivalently as
\beq
\gamma_{\sigma}(\om) = \om^{+} - \ep_{c} - \frac{V^{2}}
{\om^{+} -\ep_{f}^{*} - [\Sigma_{\sigma}(\om)-\Sigma_{\sigma}(0)]}
\eeq
where $\ep_{f}^{*}$ (Eq.(3.5)) is the renormalized level,
$\ep_{f}^{*} = \ep_{f}+\Sigma^{R}_{f}(0) \equiv
\ep_{f}+\tilde{\Sigma}^{R}_{\sigma}(0)$ (for either spin $\sigma$, as
follows directly from symmetry restoration Eq.(4.10)); and from Eq.(4.3) we
have used trivially that $\tilde{\Sigma}_{\sigma}(\om) - 
\tilde{\Sigma}_{\sigma}(0) =\Sigma_{\sigma}(\om) - \Sigma_{\sigma}(0)$.

  The particular algorithm employed is now specified, for an arbitrary conduction band $\rho_0(\ep)$. In practice we choose 
to work with specified $\ep_{c}, V^{2}, x$ and $e_{f}$, 
with the bare parameters $U$ and $\ep_{f}$ determined by solution; rather
than with $\ep_{c}, V^{2}, \ep_{f}, U$ specified and $x, e_{f}$
then determined. The two are of course entirely equivalent; we simply find the
former to be optimal in practice. So for any given $\ep_{c}, V^{2},
e_{f}$ and $x = \tfrac{1}{2}U|\mu|$, the algorithm is as follows:

(i) `Startup'. Eqs.(4.11) are first solved for the MF propagators
$\{g^{\nu}_{\sigma}(\om;e_{f},x) \}$ ($\nu = c$ or $f$), following the 
procedure specified in Eqs.(4.4-6). From this, the polarization bubble 
${^{0}\!\Pi}^{+-}(\omega)$ (and hence $\Pi^{+-}(\omega)$) follows 
directly, see Eq.(4.17). 
$\Sigma_{\sigma}(\om)$ is given by Eq.(4.18), in which the host/medium
propagator $\tilde{\cal{G}}_{-\sigma}(\om)$ is initially taken to be the
MF $g^{f}_{-\sigma}(\om)$, thus generating the `startup' 
$\Sigma_{\sigma}(\om)$.

(ii) Symmetry restoration. The $\om =0$ SR condition Eq.(4.13) is now
solved for the interaction $U$. This simply requires varying $U$ in 
Eq.(4.18) for $\Sigma_{\sigma}(\om =0)$ until Eq.(4.13) is satisfied
(the $U$-dependence of $\Sigma_{\sigma}(0)$ arising both from the trivial
$U^{2}$ prefactor in Eq.(4.18) and the explicit $U$-dependence of
$\Pi^{+-}(\om)$, see Eq.(4.17)). The local moment
follows immediately, $|\mu| = 2x/U$.

(iii) Luttinger condition. With an input guess for the renormalized level
$\ep_{f}^{*}$, and hence $\gamma_{\sigma}(\om)$ from Eq.(4.19),
Eqs.(4.5-6) are readily solved (as there described) for $G^{c}(\om)$ and
$S(\om)$; and $G^{f}(\om)$ follows directly from Eqs.(4.1,2b).
The total band filling is trivially computed from $\tfrac{1}{2}(n_{c}+n_{f})
=\int^{0}_{-\infty}d\om [D^{c}(\om)+D^{f}(\om)]$, and compared to the
Luttinger condition Eq.(3.6) (in which $\tilde{\ep}_{f}^{*} = 
\ep_{f}^{*}/V^{2}$). The renormalized level $\ep_{f}^{*}$ is then simply 
varied until Eq.(3.6) is self-consistently satisfied; the corresponding bare
$\ep_{f}$ follows directly from $\ep_{f} = \ep_{f}^{*} -
\tilde{\Sigma}^{R}_\sigma(0)$.

(iv) The resultant $S(\om)$ is then used in Eq.(4.16) to generate a new
host/medium propagator $\tilde{\cal{G}}_{-\sigma}(\om)$; and hence from
Eq.(4.18) a new $\Sigma_{\sigma}(\om)$. Now return to step (ii) and iterate
until full self-consistency is reached. 

  We find the above algorithm to be efficient, converging typically after
$\sim 6$ iterations and computationally fast on a modest PC. The outcome is a
fully self-consistent solution of the problem, with $\{\ep_{c}, V^{2},
\ep_{f},U\}$ and $e_{f},x$ all known (uniquely
so in practice). If one wishes instead to work with a specified bare parameter
set $\{\ep_{c},V^{2},\ep_{f},U\}$ one simply repeats the above
procedure, varying $e_{f}$ and $x =\tfrac{1}{2}U|\mu|$ until the desired 
$\ep_{f},U$ are obtained. The particle-hole symmetric PAM studied in
[34,35], with $\ep_{f} = -\tfrac{U}{2}$ and $\ep_{c}=0$, is a 
special case of the above algorithm; here $e_{f} = 0$ (and likewise
$\ep_{f}^{*}=0$), and the Luttinger condition is automatically satisfied
so that step (iii) above is redundant. Results arising from the fully 
self-consistent solution will be discussed in the following sections.

  Before proceeding we comment on the low-$\om$ behaviour of the transverse
spin polarization propagator $\Pi^{+-}(\om)$, given by Eq.(4.17) and 
entering the self-energy Eq.(4.18). As mentioned in \S 4.1,
this is characterised by a low-energy spin-flip scale denoted by $\om_{m}$
(and defined conveniently as the location of the maximum in Im$\Pi^{+-}(\om)$).
Such behaviour arises in all problems studied thus far within the LMA [26-35]
and has a common origin now briefly explained. If the local moment $|\mu|$ had
its pure MF value  --- {\it i.e.\ }if $|\mu|$ was determined from the
usual MF self-consistency condition (\S 4.2) $|\mu| =
|\bar{\mu}(e_{f},x=\tfrac{1}{2}U|\mu|)|$ with $|\bar{\mu}(e_{f},x)|$ 
given generally by Eq.(4.12a) --- then it is straightforwardly shown 
(following {\it e.g.\ }[27,34])
that $\Pi^{+-}(\om)$ given by Eq.(4.17) contains a pole at $\om =0$ 
identically.  In physical terms
this reflects simply the fact that the {\em pure } MF state is, locally,
a symmetry broken degenerate doublet, with zero energy cost to flip an
$f$-level spin. This is
correct only in the `free lattice' limit of vanishing hybridization $V$ 
where the $f$-levels decouple from the conduction band, resulting in a 
degenerate local moment state (a limit that we note is recovered 
exactly by the LMA, non-trivially so from the perspective of 
conventional perturbative approaches to the PAM). 
Such `cost free' spin-flip physics is not of course correct for the 
Fermi liquid phase of the PAM that is adiabatically connected to the
non-interacting limit.  But neither does it occur in this case, for 
the existence of an $\om =0$ 
spin-flip pole is readily shown to be specific {\em solely } to the 
pure MF level of self-consistency 
({\it i.e.\ }arises only if $|\mu|$ has its pure MF value specified above). 
The key point is that within the LMA the local moment $|\mu|$ is 
determined from the
symmetry restoration condition Eq.(4.10) (as in step (ii) above), which itself
reflects adiabatic continuity (see \S 4.1). In consequence Im$\Pi^{+-}(\om)$
contains not an $\om =0$ spin-flip pole, but rather a low-energy resonance
centred on a non-zero frequency $\om_{m}$, whose physical content in setting
the timescale for symmetry restoration has already been noted in \S 4.1
(and which is equivalently the low-energy Kondo lattice scale 
$\om_{m} \propto \om_{L} =ZV^{2}$ (Eq.(3.10))). 

\subsection{Single-impurity model }

  For sufficiently low energies and/or temperatures the behaviour of the
PAM is that of a coherent Fermi liquid [1,2], reflecting the lattice
periodicity and embodied in the lattice coherence scale $\omega_{L}$.
However with increasing energy and/or temperature, it has of course long 
been known
that a crossover should occur to an incoherent regime of effective 
single-impurity physics [1,2]. For that reason it is traditional 
in studies of the PAM/KLM to compare to corresponding
results for the Anderson impurity (or Kondo) model; in which only a 
single $f$-level is coupled to the host conduction band, but otherwise 
with the same bare parameters
as the PAM itself, ($\ep_{c}, V^{2}, \ep_{f}, U$). 
Here we comment
briefly on the LMA for the relevant Anderson impurity model (AIM) itself [26,27],
which will be used in \S5.

  The local impurity Green function for the AIM, which we continue to denote
by $G^{f}(\om)$, is given by $G^{f}(\om) = \tfrac{1}{2}\sum_{\sigma}G^{f}_{\sigma}(\om)$
( {\it cf } Eq.(4.1)) where
\beq
G^{f}_{\sigma}(\om) = [\om^{+} - \ep_{f} -
\tilde{\Sigma}_{\sigma}(\om)-\Delta(\om) ]^{-1}.
\eeq
One-electron coupling between the impurity $f$-level and the host is as usual
embodied in the hybridization function $\Delta(\om) = \Delta_{R}(\om) -
\ii\,\sgn(\om)\Delta_{I}(\om)$, and is given simply by
$\Delta(\om) = V^{2}g^{c}_{0}(\om)$;
where $g^{c}_{0}(\om)$ is the free lattice ($V=0$) conduction electron
propagator (Eq.(2.4)) with corresponding dos $d^{c}_{0}(\om) =
\rho_{0}(\om - \ep_{c})$
(and $\rho_{0}(\om)$ given {\it e.g.\ }for the hypercubic and Bethe lattices by 
Eqs.(2.5)). Note then that the AIM hybridization is both $\om$-dependent and,
for generic $\ep_{c} \neq 0$, asymmetric in $\om$ about the Fermi level
$\om =0$ (in contrast {\it e.g.\ }to the commonly considered wide flat-band
AIM [2] for which $\Delta_{R}(\om) =0$ and $\Delta_{I}(\om) =$ constant). 
This $\om$-dependence in $\Delta(\om)$
will of course be apparent in AIM single-particle dynamics on 
non-universal energy
scales (as seen {\it e.g.\ }in Fig. 5 below). But in the strong coupling 
Kondo regime of the AIM, characterised by an exponentially small Kondo 
scale $\om_{K}
\rightarrow 0$, universal scaling behaviour of dynamics arises in terms of 
$\omega/\omega_{K}$ (see {\it e.g.\ }[26-28,33]). In this regime the $\om$-dependence
of $\Delta(\om)$ is naturally immaterial, and only $\Delta(\om =0)$ is relevant.
With that in mind, for later use we denote
by $\Delta_{0}=\Delta_{I}(\om =0)$ the local hybridization strength at the Fermi level,
\beq
\Delta_{0} = \pi V^{2}\rho_{0}(-\ep_{c}).
\eeq

The self-energies $\{\tilde{\Sigma}_{\sigma}(\om)\}$ for the AIM in Eq.(4.20)
are again given by Eq.(4.3); with
$\bar{n}$ and $|\bar{\mu}|$ by Eq.(4.12), where now the
MF spectral densities $d^{f}_{\sigma}(\om) \equiv d^{f}_{\sigma}(\om;e_{f},x)$
naturally pertain to the MF propagators for the AIM, given by ({\it cf }
Eq.(4.11b)) $g^{f}_{\sigma}(\om) = [\om^{+} -e_{f} +\sigma x - 
\Delta(\om)]^{-1}$.
The dynamical contributions to the self-energies, $\{\Sigma_{\sigma}(\om) \}$ 
(Eq.(4.3)) are likewise given [26,27] by Eqs.(4.15) or (4.18),
with the self-consistent host/medium propagator 
$\tilde{\cal{G}}_{-\sigma}(\om)$ 
appropriate to the PAM now replaced simply by the AIM propagator 
$g^{f}_{-\sigma}(\om)$. And the symmetry restoration condition for the AIM is
again given by Eq.(4.9) [27], whence $\Sigma_{f}^{R}(0) =
\tilde{\Sigma}^{R}_{\sigma}(0)$ (for either spin $\sigma$) where 
$\Sigma_{f}(\om)$ here denotes the AIM single self-energy. Finally, 
with $\ep_{imp}^{*} = \ep_{f} + \Sigma^{R}_{f}(0)$ denoting the impurity
renormalized level (as for the PAM, Eq.(3.5)), the Luttinger
integral theorem $I_{L} =0$ (Eq.(3.1)) yields directly the Freidel sum rule for
the AIM [2,42]: $\ep_{imp}^{*} + \Delta_{R}(0) =
\Delta_{0}$tan$(\tfrac{\pi}{2}n_{imp})$,
with $n_{imp}$ as usual the excess charge induced by addition of the impurity,
and which is the AIM analogue of Eq.(3.6) appropriate to the PAM.
The LMA for the single-impurity model is readily implemented, as detailed in
[27], with both symmetry restoration and the Luttinger theorem satisfied.

\section{Results }

 We turn now to results arising from the LMA specified above. Following
consideration of dynamics on all energy scales (\S 5.1), the dependence of the
coherence scale $\om_{L}$ on bare/material parameters is
obtained in \S 5.2 and compared to corresponding results for the Kondo scale 
$\om_{K}$ of the single-impurity Anderson model. The central issues 
are considered in \S 5.3: the $\omega/\omega_{L}$-scaling behaviour of
single-particle spectra in the strong coupling Kondo lattice regime, and
their evolution from the low-energy physics characteristic of the coherent
Fermi liquid through to the emergence at high energies of single-impurity 
scaling behaviour. Finally, \S 5.4 discusses our results in the context 
of Nozi\e res'
problem of `exhaustion' [37,38] and recent work on that issue.

\subsection{All scales overview }
\begin{figure*}[t]
\begin{center}
\includegraphics*[scale=0.60]{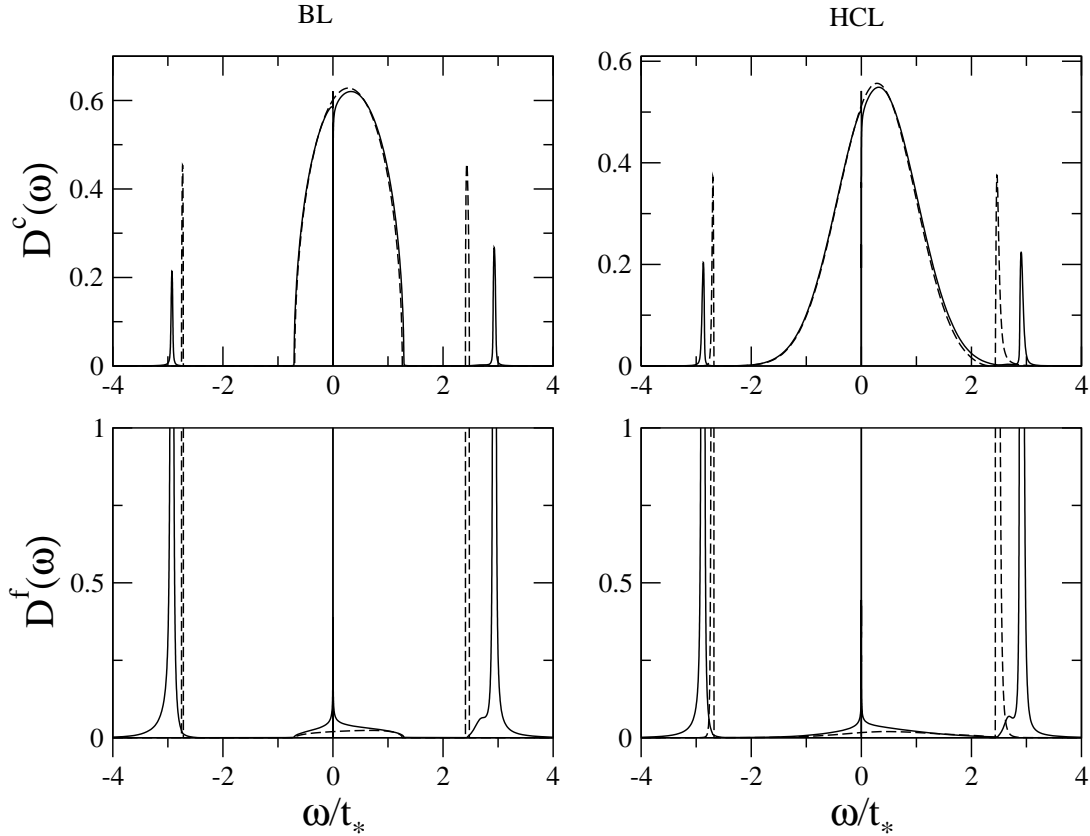}
\end{center}
\protect\caption{
All scales view of LMA $c$- and $f$-electron spectra
(solid lines): $D^{c}(\om)$ and $D^{f}(\om)$ {\it vs } $\om$
$(\equiv \om/t_{*})$, for Bethe lattice (left panels) and hypercubic
lattice (right panels). For $U \simeq 5.1$  $(x=\tfrac{1}{2}U|\mu| = 2.5)$,
$V^{2} =0.2$, $\ep_{c} = 0.3$ and $\eta = 1-2|\ep_{f}|/U =0$.
Corresponding MF spectra are also shown (dashed lines).
}
\end{figure*}

     For obvious physical reasons the 
primary interest in the PAM resides
both in the strongly correlated Kondo lattice regime, and on energies on the 
order of the coherence scale $\om_{L} =ZV^{2}$ and 
(essentially arbitrary)
multiples thereof. We begin however with an overview on all energy scales
--- encompassing `band scales' $\om \sim {\cal{O}}(1)$
($1 \equiv t_{*})$ and energies $\omega \sim \epsilon_{f}$ or $\epsilon_{f}+U$
characteristic of the $f$-electron Hubbard satellites. 
In contrast to the low-energy sector, dynamics here will naturally be 
non-universal: dependent on  essentially all bare material/model 
parameters, and lattice specific. 
An overview is nonetheless instructive, showing clearly the roles of asymmetry
(in both the conduction band and $f$-levels), and of the lattice type, 
as well as qualitative effects of depleting the conduction band filling. 
In addition, it enables broad comparison both to dynamics arising
at the crude level of pure MF (\S 4.2) and to corresponding results for the 
Anderson impurity model ($\S 4.5$).

  Figs.2 and 3 show spectra typical of metallic heavy
fermion behaviour in strong coupling:
$U \simeq 5.1$ ($x=\tfrac{1}{2}U|\mu|\\ = 2.5$), $V^{2} =0.2$ and 
$\ep_{c} = 0.3$. In Fig.2, $\eta = 1 + 2\ep_{f}/U =0$ is taken --- 
corresponding to symmetric $f$-levels $\ep_{f} = -\tfrac{U}{2}$,  
but with asymmetry in the conduction band ($\ep_{c} \neq 0$). To illustrate 
the effects of the lattice, the $c$- and $f$-electron spectra are 
each shown for 
both the hypercubic lattice (HCL) and Bethe lattice (BL). In either case the 
$f$-level charge $n_{f} \simeq 0.99$. The conduction band fillings,
likewise determined from spectral integration, 
differ little for the two lattices, $n_{c} \simeq 0.64$ (BL) and $0.69$ (HCL)
(each being within $\sim 2\%$ of the asymptotic strong coupling result 
Eq.(3.13) for $n_{c}$).

  The overwhelming intensity of the $f$-spectra shown in Fig.2 is naturally in 
the Hubbard satellites, well separated from the band scales 
$\omega \sim {\cal{O}}(1)$ (and in consequence sharply distributed). Their peak 
maxima are symmetrically positioned about the Fermi level --- reflecting the
fact that the $f$-levels themselves are symmetric ($\eta =0$) --- and largely 
unaffected by the presence of asymmetry in the conduction band. The most 
important 
feature of the $f$-spectra is of course the well known many-body resonance
at low energies. Its rich structure, considered in detail in \S 5.3, 
is naturally not resolved here.
What is however evident from Fig.2 is the relative unimportance of the 
host lattice in determining the $f$-spectra on the all scales level 
shown. This is in contrast
to the local conduction electron spectra (top panels, Fig.2). Here, aside from
weakly remnant Hubbard satellites whose intensity diminishes steadily 
with increasing
$U$, the $c$-spectra are clearly dominated by the asymmetrically distributed 
envelope of the free ($V=0$) conduction 
band spectrum, semi-elliptic for the BL and Gaussian for the HCL. As 
mentioned at the end of \S 3 this is physically natural, reflecting 
that in the strongly correlated regime the conduction band is very 
weakly coupled to the $f$-levels; albeit that such coupling 
is of course the key feature of the problem on low energy scales, 
where it leads to 
many-body structure in the $D^{c}(\om)$ (again barely visible on the scales 
shown).
  In Fig.3, shown for the HCL with $\eta =0.4$, there is now particle-hole 
  asymmetry in the $f$-levels as well as in the conduction band; producing
the additional spectral signature of asymmetry in the positions of the Hubbard 
satellites, but otherwise little change in broad terms.
\begin{figure*}[t]
\begin{center}
\includegraphics*[scale=0.60]{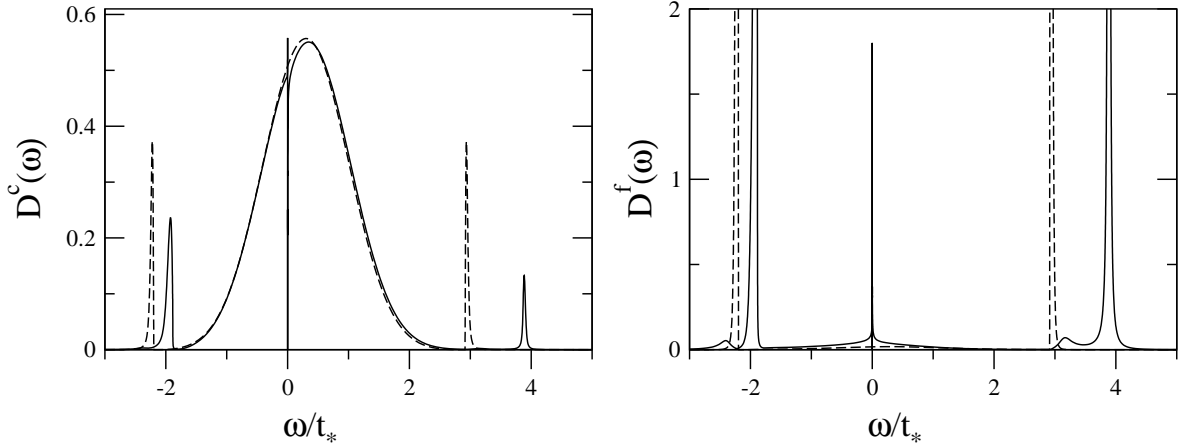}
\end{center}
\protect\caption{
Local $c$- and $f$-electron spectra for the HCL: $D^{c}(\om)$
and $D^{f}(\om)$ {\it vs } $\om$, shown for $\eta = 1-2|\ep_{f}|/U =0.4$
with remaining parameters as in Fig.2. Dashed lines: corresponding MF spectra.
}
\end{figure*}

  Figs.2 and 3 also show direct comparison to the corresponding spectra at pure 
mean-field level (dashed lines). At first sight, and on the all scales level 
shown, these appear to provide a reasonable first approximation to dynamics. 
That is {\it not} of course the case on the all important low-energy scales 
that dominate the physics of the PAM in strong coupling: MF clearly lacks 
any hint of the many-body resonance
in the $f$-spectra and its counterpart in $D^{c}(\om)$ --- unsurprisingly
given the absence of correlated electron dynamics at this crude level --- and
in fact
for $D^{f}(\om)$ is seen to be qualitatively deficient for essentially all
$|\om| \lesssim 1$. For the $c$-electron spectra however, and again excepting
the lowest energies, MF is qualitatively reasonable, reducing in strong 
coupling to precisely the free conduction band spectrum (as is readily 
shown directly from Eqs.(4.11)). In addition, MF also captures qualitatively 
the dominant Hubbard satellites in the $f$-electron spectra; albeit that 
many-body broadening effects, arising from the spin-flip dynamics included 
in the LMA, lead both to a broadening
and slight shift of the satellites (which can be understood quantitatively 
in terms of the $\om$-dependence of the dynamical self-energies 
$\Sigma_{\sigma}(\om)$, although we do not pursue that further here).

  We consider now the qualitative effect of depleting the conduction
band filling $n_{c}$, obtained by increasing $\ep_{c}$ significantly.
Fig.4 shows HCL spectra for $\ep_{c} =1.5$, $U\simeq 7$ 
($x=3.5$), $\eta =0.8$ and $V^{2} = 1.25$; for which the resultant
$n_{c} \simeq 0.2$ (and $n_{f} \simeq 0.85$). Save for the large 
$\epsilon_{c}$ the remaining parameters
have no special significance, and the large $V^{2}$ has simply been 
chosen so that the resultant low-energy scale $\om_{L}$ (discussed in 
detail in \S 5.2) is not 
so small as to be in effect invisible in the figure shown. Depleting 
$n_{c}$ in this way has a marked effect on the conduction electron spectra. 
In contrast to Figs.2,3 for $\ep_{c}=0.3$ --- where the low-energy
`antiresonance' in $D^{c}(\om)$ is carved out of the free conduction band 
envelope --- Fig.4 shows that the low-$n_{c}$  conduction spectrum now 
contains a sharp low-energy resonance
lying in the tail of the free conduction band envelope,
akin to that appearing ubiquitously in the local $f$-spectra. 
The essential origin of the resonance is readily seen from the 
quasiparticle behaviour discussed in \S 3: from the 
quasiparticle $D^{c}(\om)$ Eq.(3.11a), the Fermi level
$D^{c}(\om =0) \sim \rho_{0}(-\ep_{c} + 1/\tilde{\ep}_{f}^{*})$
(with $\tilde{\ep}^{*}_{f} = \ep_{f}^{*}/V^{2}$ and $\ep_{f}^{*}$
the renormalized level); while for large $\ep_{c} =|\ep_{c}|$, Eq.(3.12) 
shows that 
$-\ep_{c} + 1/\tilde{\ep}^{*}_{f} \rightarrow 0$ as $n_{f} \rightarrow 1$.
In consequence, $D^{c}(\om =0) \sim \rho_{0}(\om =0)$ is effectively 
pinned at the
Fermi level to its free lattice limit (with $\rho_{0}(0) = 1/\sqrt{\pi}$
for the HCL marked explicitly on Fig.4); and the width of the resultant
resonance is ${\cal{O}}(\om_{L})$, as again follows from the quasiparticle
$D^{c}(\om)$, Eq.(3.11a). The  low-energy resonance arising for low $n_{c}$ 
in the $c$-electron spectrum thus reflects directly the Fermi liquid nature of 
the ground state.

\begin{figure}[h]
\begin{center}
\includegraphics[scale=0.32]{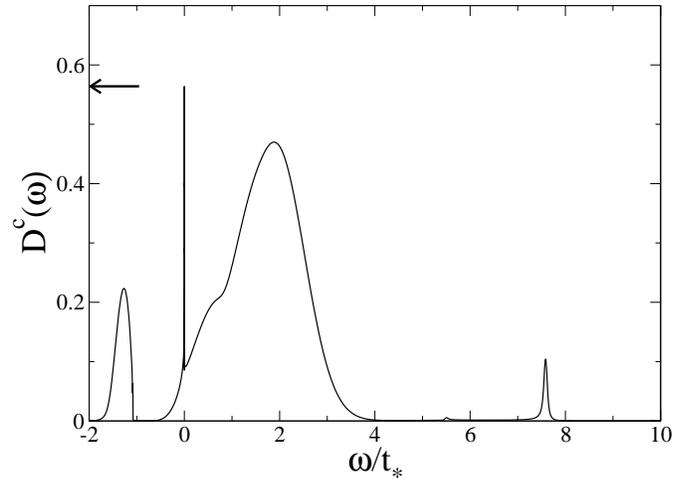}
\end{center}
\protect\caption{
Depleting the conduction band: $D^{c}(\om)$ {\it vs } $\om$
for the HCL with $\ep_{c} =1.5$, $U \simeq 7$ ($x=3.5$),
$\eta =0.8$ and $V^{2} =1.25$, corresponding to $n_{c} \simeq 0.2$. 
The conduction
band spectrum now shows a sharp resonance in the vicinity of the Fermi level
$\om=0$ (cf Figs.2,3) as explained in text; $\rho_{0}(\om=0) = 1/\sqrt{\pi}$
is indicated by an arrow.
}
\end{figure}
  In Fig.5, comparison is made between an $f$-electron spectrum for the 
PAM for the HCL (solid line) and its counterpart for the single-impurity 
Anderson model (dash-ed line); the bare parameters chosen for illustration
being $\eta =0$ (symmetric $f$-level(s)), $U\simeq 1.3$ 
($x=0.5$) and $V^{2} =0.2$, $\ep_{c} =0.3$ (corresponding to a hybridization 
strength, Eq.(4.21), of $\pi \Delta_{0} = (\pi V)^{2}\rho_{0}(-\ep_{c}) 
\simeq 1$).
The spectra, specifically $\pi\Delta_{0}D^{f}(\om)$, are compared on the 
all scales level in the main figure ({\it vs}
$\om \equiv \om/t_{*}$); and on the low-energy resonance scale in the inset.
The main figure shows in addition
the impurity spectrum for $\ep_{c}=0$ (with the same parameters 
otherwise), which is of course the fully particle-hole symmetric AIM. 
The first point to note here is 
obvious: excepting the low-energy sector, the PAM and corresponding 
AIM spectra are qualitatively very similar on the `all scales' level 
(which is not specific to the particular
parameter set employed). This general characteristic is in agreement 
with results from a numerical renormalization group (NRG) study of the 
PAM and AIM [8], see {\it e.g.\ }Figs.1,4,6 of [8]. Note further, in 
comparison to the fully particle-hole symmetric
AIM (dotted line), that for the moderate value of $U$ chosen in Fig.5 the 
asymmetry in the conduction band ($\ep_{c} \neq 0$) shows up weakly in the 
Hubbard satellites, both in their intensities and maxima (which are not quite 
symmetrically positioned about the Fermi level).

  It is naturally in the low-energy behaviour that the PAM and AIM 
spectra differ significantly, as evident in the inset to Fig.5. In 
particular, for the PAM the lattice coherence generates a pseudogap 
above the Fermi level, as indeed expected
from the quasiparticle behaviour of $D^{f}(\om)$, Eq.(3.11b)
(albeit that the relative weakness of the pseudogap here reflects 
the moderate $U$ considered, see \S 5.3). However even at low energies 
one does not learn much from comparison of PAM and AIM dynamics on an 
{\it absolute } scale, {\it e.g.\ vs }
$\om \equiv \om/t_{*}$, and for a given set of bare parameters 
$\{\ep_{c}, V^{2}, U, \eta \}$. For the parameters chosen in Fig.5, it
so happens that the low-energy scales for the PAM and AIM are very similar 
(with quasiparticle weights $Z \approx 0.1$). But that is not generically
so: as discussed in \S 5.2 the PAM lattice coherence scale and the Kondo scale 
for the AIM will in general be quite different for given bare 
parameters [8,23] (they 
are after all physically distinct models), whence comparison of the two on
an absolute scale is barely informative. 
What is required by contrast --- particularly in strong coupling where
there is a pristine separation between asymptotically vanishing low-energy
scale(s) and non-universal scales such as $\Delta_{0}$, $t_{*} \equiv 1$ or
$U$ --- is comparison of the universal {\it scaling behaviour } of the two 
models; in which the dependence of the respective low-energy scales 
on bare parameters is thereby eliminated and the underlying scaling 
behaviour exposed.  This we believe is the most convincing (and possibly 
only) way to establish a connection between the {\it high-energy scaling} 
behaviour of the PAM/KLM and underlying single-impurity physics. 
That key issue is considered in \S 5.3. 
\begin{figure}[h]
\begin{center}
\includegraphics[scale=0.32]{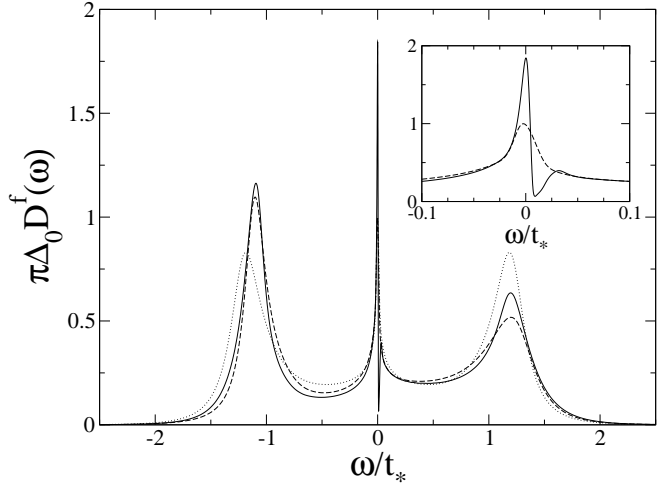}
\end{center}
\protect\caption{
All scales comparison of local $f$-electron spectra for the PAM and
Anderson impurity model:
$\pi\Delta_0 D^f(\om)$ {\it vs } $\om \equiv \om/t_{*}$
(with $\Delta_{0} = \pi V^{2}\rho_{0}(-\ep_{c})$); for $\ep_c=0.3$,
$U \simeq 1.3 (x=0.5), V^{2}=0.2$ and $\eta =0$. Solid line: PAM.
Dashed line: AIM. Inset: comparison on the resonance scales. The 
impurity spectrum for $\ep_{c}=0$ is also shown on the main 
figure (dotted line, the particle-hole symmetric AIM).
}
\end{figure}

\subsection{Low-energy scale }

  We first consider briefly how the low-energy coherence scale for the PAM, 
$\om_{L}=ZV^{2}$ (Eq.(3.10)), is found within the LMA to depend on the 
bare/material parameters $(\ep_{c},\\V^{2}, U,\eta)$ in the strong coupling 
Kondo lattice regime where $n_{f} \rightarrow 1$; and how it compares to the 
Kondo scale for the corresponding AIM, $\om_{K} \equiv Z_{imp}V^{2}$ (with 
$Z_{imp}$ denoting the quasiparticle weight for the single-impurity model). 
  The scales $\om_{L}$ and $\om_{K}$ are indeed found to be exponentially
small in strong coupling (as opposed to algebraically small, such as 
arises using perturbation theory in $U$ or variants thereof such as 
modified (iterated) perturbation theory [14-17]); leading thereby to the 
clean scale-separation that is a prerequisite to the scaling considerations 
of \S 5.3.  The essential findings here, discussed below, agree with the 
NRG study of [8] and results obtained from the large-$N$/slave boson 
mean-field (SBMF) approximation [23]:
(a) That the scales $\om_{L}$ and $\om_{K}$
are found to have the same exponential dependence on the bare parameters, but
(b) depend very differently on the conduction band filling $n_{c}$; the lattice
scale $\om_{L}$ being enhanced relative to its single-impurity counterpart 
as $n_{c} \rightarrow 1$, but strongly diminished for 
low $n_{c}$.

  The material dependence of the AIM Kondo scale $\om_{K} (\propto
\om_{m})$ arising
within the LMA can be obtained analytically in strong coupling
by direct analysis of the symmetry
restoration condition Eq.(4.13). That was considered explicitly in [27] for the
case of a general impurity (with $f$-level asymmetry embodied as usual
in $\eta \equiv 1 -2\tfrac{|\ep_{f}|}{U}$), but a symmetric host band. 
It is straightforward to extend the analysis of [27] to include the
$\om$-dependence of the hybridization $\Delta(\om)$ arising from an 
asymmetric host conduction band (which as anticipated in \S 4.5 does not
affect the final answer). Noting that the AIM  hybridization strength 
is given by $\Delta_{0} = \pi V^{2}\rho_{0}(-\ep_{c})$
(Eq.(4.21)), this yields 
\beq
\om_{K} \propto ~
\exp\left( -\frac{U}{8V^{2}\rho_{0}(-\ep_{c})}\frac{(1-\eta^{2})}
{f(\eta^{2})}\right)
\eeq
(with the proportionality determined simply by a high-energy cutoff [27]); where
$f(\eta^{2}) =
\tfrac{1}{2}[1+ \sqrt{(1-\eta^{2})}]$ ($ \in [1,\tfrac{1}{2}]$ for asymmetries
$|\eta| \in [0,1]$ relevant to the Kondo regime). 
The exchange coupling $J$ for the corresponding Kondo model, obtained from 
the AIM in strong coupling by a Schrieffer-Wolff transformation [2], is
given by $J = V^{2}[\tfrac{1}{|\ep_{f}|} + \tfrac{1}{U-|\ep_{f}|}]
\equiv 4V^{2}/[U(1-\eta^{2})]$; whence Eq.(5.1) is equivalently 
$\om_{K} \propto  \exp(-1/[2\rho_{0}(-\ep_{c})J f])$.
As pointed out in [27] the exponent here differs in general by the factor 
$f(\eta^{2})$
from the exact result for the Kondo model,
being as such exact only for the symmetric case where $\eta =0$ (although 
note that $f$ is slowly varying in $\eta$, lying {\it e.g.\ }within
10$\%$ of unity for $\eta <0.6$). That notwithstanding we regard recovery of
an exponentially small Kondo scale, close to the exact result in an obvious
sense, as non-trivial; and add that provided the scale is indeed exponentially
small, so that a clean separation of low (universal) and non-universal energies
arises, its precise dependence on the bare parameters is in essence irrelevant
to the issue of scaling in terms of $\om/\om_{K}$ (as seen in [27] for the AIM 
itself).

\begin{figure}[h]
\begin{center}
\includegraphics[scale=0.32]{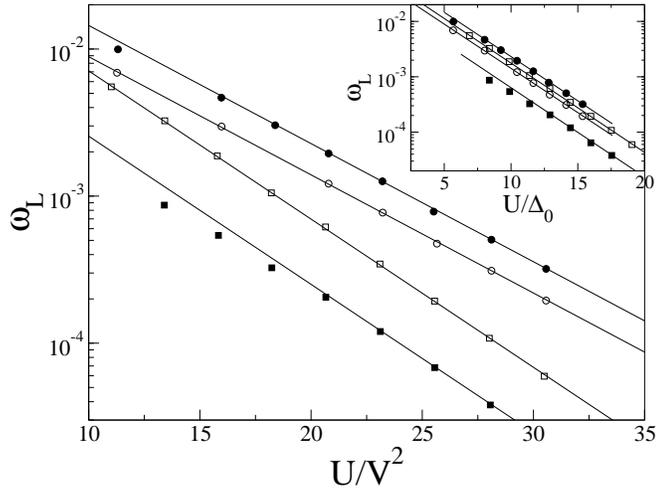}
\end{center}
\protect\caption{
 PAM coherence scale $\om_{L}$ on a log-scale {\it vs } $U/V^2$;
for the BL with $\eta =0$, $\ep_c =0.1$ (solid circles) and
$\ep_c =0.6$ (solid squares). Comparison is made to corresponding results
for the AIM Kondo scale $\om_{K}$: $\ep_{c} =0.1$ (open circles) and
$\ep_{c} =0.6$ (open squares). $\om_{L}$ is enhanced relative to
$\om_{K}$ for $\ep_{c} = 0.1$ but diminished for $\ep_{c} =0.6$
(see also Fig.7).  The straight lines indicate the exponents 
given by Eq.(5.1), found for both models.
Inset: same results now shown {\it vs } $U/\Delta_{0}$ (with $\Delta_{0}=\pi
V^{2}\rho_{0}(-\ep_{c})$). The strong coupling gradients are then 
common for different $\ep_{c}$, as implied by Eq.(5.1).
}
\end{figure}

  For the PAM the coherence scale $\om_{L}$ is likewise obtained from solution 
of the symmetry restoration condition Eq.(4.13), in this case numerically 
following the procedure detailed in \S 4.4 (and with $\om_{L}$ found to be 
proportional to the spin-flip scale $\om_{m}$ as noted in \S 4.4). In the 
strong coupling Kondo lattice regime $\om_{L}$ is again
found to be exponentially small, with its exponential dependence
on the bare parameters the same as $\om_{K}$ for the AIM. This is illustrated
in Fig.6 (for the BL) where, with $\eta =0$ and for $\ep_{c} = 0.1$ and $0.6$, 
the resultant $\om_{L}$ is plotted on a logarithmic scale {\it vs}  $U/V^{2}$; 
and compared to the counterpart $\om_{K}$ results for the AIM itself. 
For given $\ep_{c}$ the asymptotic PAM and AIM curves are parallel, indeed 
indicating common $U/V^{2}$-dependence for the exponents of the two scales.
When plotted {\it vs}  $U/V^{2}$ as in the main figure, the slopes for 
different $\ep_{c}$ clearly differ; but when shown {\it vs }
$U/\Delta_{0} = U/(\pi V^{2}\rho_{0}(-\epsilon_{c}))$
as in the inset to Fig.6, the gradients for different $\epsilon_{c}$ are now 
common in strong coupling, as implied by the exponential 
dependence of Eq.(5.1). 
The dependence of the exponents on the $f$-level asymmetry, as in Eq.(5.1), may 
likewise be verified by varying $\eta$. And the same exponential dependence, Eq.(5.1),
is found whether the BL or HCL is considered.
\begin{figure}[h]
\begin{center}
\includegraphics[scale=0.32]{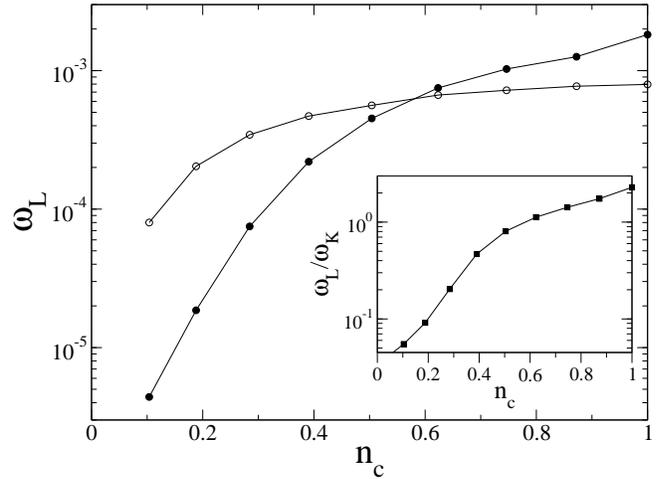}
\end{center}
\protect\caption{
PAM coherence scale $\om_{L}$ {\it vs } $n_{c}$, for the BL with
$U/V^{2} =23$ and $\eta =0$ (solid circles), compared to the AIM Kondo
scale $\om_{K}$ (open circles). Inset: $\om_{L}/\om_{K}$ {\it vs }
$n_{c}$.
}
\end{figure}

\begin{figure*}[t]
\begin{center}
\includegraphics*[scale=0.50]{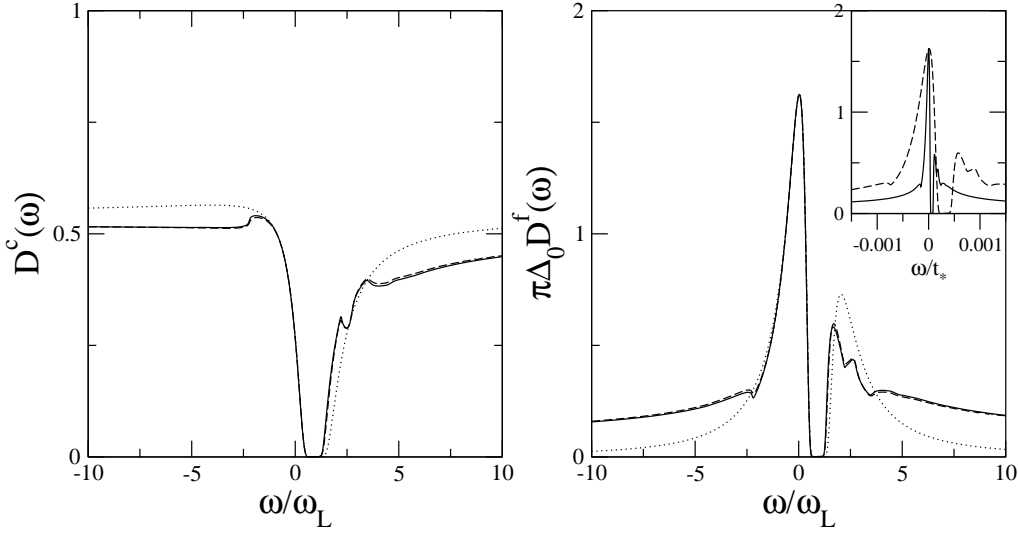}
\end{center}
\protect\caption{
 Universal spectral scaling. $D^{c}(\om)$ and
$\pi\Delta_{0}D^{f}(\om)$ {\it vs } $\om^{\prime} =\om/\om_{L}$ for the
HCL with $\ep_{c}=0.2$ and $\eta =0$. The spectra for $U \simeq 5.1$ (dashed
line) and $U \simeq 6.6$ (solid line) collapse to a common scaling form 
as a function
of $\om/\om_{L}$. Inset: corresponding $f$-spectra shown by contrast on an
absolute scale, {\it vs } $\omega/t_{*}$. The main figures also show the
asymptotic low-$|\om^{\prime}|$ quasiparticle spectra (Eqs.(3.11)),
dotted lines;
to which the full scaling spectra reduce for $|\om^{\prime}| \lesssim 1$.
}
\end{figure*}
  While the {\it exponents } of the scales $\om_{L}$ and $\om_{K}$ have
 the same dependence on bare parameters, their dependence on the conduction 
 band filling $n_{c}$ (which itself is determined solely by $\ep_{c}$ in
 strong coupling, see Eq.(3.13)) is quite distinct for the two models.
 This is evident already in Fig.6 but seen more clearly in Fig.7 where, 
 for $U/V^{2} = 23$ (and $\eta =0$) we show $\om_{L}$ and $\om_{K}$ {\it vs}
 $n_{c}$ (their ratio being shown in the inset). For $n_{c} \rightarrow 1$,
$\om_{L}/\om_{K} = F(n_{c}) > 1$ and the lattice scale is enhanced over
its AIM counterpart; while with decreasing $\ep_{c}$ (and hence $n_{c}$)
$\om_{L}$ diminishes progressively in comparison to $\om_{K}$, such that
$F(n_{c}) \rightarrow 0$ as $n_{c} \rightarrow 0$. As noted above this 
general behaviour is in agreement with NRG [8] and SBMF results [23]. 
It is by contrast quite distinct 
from results arising from a Gutzwiller variational treatment [24] in 
which the lattice scale is always enhanced over its AIM counterpart; 
or from approaches based on lattice extensions of the non-crossing 
approximation (NCA) [18,19] in which the lattice scale, while in general 
moderately enhanced compared to $\omega_{K}$, is essentially equivalent 
to the AIM scale.

\subsection{Scaling }

  The preceding discussion of how $\om_{L}$ and $\om_{K}$ for the two 
different models depend on bare parameters has been included in part 
because of the interest it has hitherto attracted in the literature.
As noted earlier however we regard this matter as quite subsidiary in 
comparison to the strong coupling scaling behaviour {of the lattice model 
itself}, as now considered.

  First we show that in the Kondo lattice regime the LMA indeed
leads to universal scaling of single-particle dynamics in terms of 
$\om/\om_{L}$, for fixed $\ep_{c}$ and $\eta$. This is illustrated
for the hypercubic lattice in Fig.8, for $\ep_{c} =0.2$ and $\eta =0$. The 
inset to the figure shows the $f$-electron spectrum $\pi \Delta_{0}D^{f}(\om)$ 
on an absolute scale, {\it i.e.\ vs}  $\om$ ($\equiv \om/t_{*}$), for
$V^{2} =0.2$ and two different interaction strengths, 
$U \simeq 5.1$ ($x=2.5$) and $U\simeq 6.6$
($x=3.25$). In either case the resultant conduction band filling 
$n_{c} \simeq 0.78$, in agreement with the asymptotic result Eq.(3.13) 
which shows that $n_{c}$ is determined by $\ep_{c}$ alone; and likewise 
$n_{f} \simeq 0.99$. As seen from the inset the two spectra are quite 
distinct on an absolute scale and dependent on the bare model parameters,
reflecting the exponential diminution of the coherence scale $\omega_{L}$ 
with increasing $U/V^{2}$ as in \S 5.2 above. The main part of Fig.8 
by contrast shows both $\pi\Delta_{0}D^{f}(\om)$ and the corresponding 
conduction spectrum $D^{c}(\om)$, now {\it vs} $\om^{\prime} =\om/\om_{L}$,
from which collapse to common scaling forms and hence universality is 
clear. While scaling has been demonstrated here by considering 
fixed $V^{2}$ upon increasing $U$ 
in the KL regime, it is as expected dependent solely on the ratio $U/V^{2}$ 
(the same scaling spectrum arising for fixed $U$ upon decreasing $V^{2}$).

  As discussed in \S 3, adiabatic continuity to the non-interacting limit
requires that for sufficiently low $\om^{\prime}$ the scaling
spectra should reduce to the quasiparticle forms Eq.(3.11).
That this behaviour is indeed recovered correctly by the LMA is also 
seen in Fig.8, 
where the resultant quasiparticle spectra are shown for comparison 
(dotted lines, as given by Eq.(3.11) with $\tilde{\ep}^{*}_{f} \equiv
\tilde{\ep}^{*}_{f}(\ep_{c})$ obtained from Eq.(3.12) with $n_{f} =1$): 
in the vicinity of the Fermi level ($\om^{\prime} =0$), and up 
to $|\om^{\prime}| \simeq 1$ or so, agreement with the quasiparticle behaviour 
is essentially perfect. For larger $|\om^{\prime}|$ by contrast, an evident 
departure from this simple low-$\om^{\prime}$ asymptotic
behaviour sets in; in particular, the quasiparticle 
$f$-electron spectra for large $|\om^{\prime}|$ are seen to
decay much more rapidly ($\sim 1/|\om^{\prime}|^{2}$) than the 
full LMA results, which show instead slowly decaying spectral tails. 
The latter, which as shown below decay logarithmically slowly, 
are a key feature of dynamics
(see Figs.11-13); reflecting genuine many-body scattering/lifetime
effects, setting in for $|\om^\prime| \gtrsim 1-10$ and dominating the
scaling spectra (as well as transport properties on corresponding 
temperature scales, see {\it e.g.\ }[34,35]). 
Here we note in passing that scaling spectra arising from 
a SBMF approximation are just
the quasiparticle forms themselves, and are evidently deficient
except for the lowest energy scales; and similarly
that dynamics arising from modified (iterated) perturbation theory [16,17] 
amount to little more than quasiparticle form, and similarly lack 
non-trivial high-energy scaling behaviour [35].
 We also add that the spectral substructure seen
in Fig.8 just above the upper edge of the gap is not a numerical artefact,
and using the LMA self-energies
can in fact be understood physically in terms of correlated `strings' of
$f$-electron spin-flips on distinct lattice sites.
It is however destroyed thermally on temperature scales which are a
small fraction of $\om_{L}$ itself (as will be shown in subsequent work),
and as such is but a minor feature of dynamics that we do not pursue further 
here.

The spectra shown in Fig.8 display an evident gap lying slightly above 
the Fermi level (strictly a pseudogap for the HCL), as found also in 
approaches based on lattice extensions of the NCA [18-20]. 
That such behaviour arises is to be expected, for it occurs likewise
in the quasiparticle spectra Eqs.(3.11) (see also the discussion at the end
of \S 2.1). Note however that this gap become `fully developed' only 
in the strong coupling Kondo lattice regime; for weaker interaction strengths
outside the scaling spectrum it is by contrast incompletely formed 
and evident only as a weaker pseudogap, as seen clearly
{\it e.g.\ }in Fig.5. But for sufficiently strong coupling we find
that a well developed gap always arises (as the quasiparticle forms
would suggest). Such behaviour is also found in recent NRG calculations [8]
for $n_c \sim 1$, but not for significantly lower conduction band
fillings -- see {\it e.g.\ }Fig.7 of [8] for $D^c(\om)$ with
$n_c=0.6$ where, by contrast, only weaker pseudogap behaviour
is evident. However the spectrum {\it e.g.\ }in Fig.7 of [8] is clearly
not close to strong coupling behaviour; as evidenced both from the fact 
that the $D^c(\om)$ shown there departs significantly from the free conduction
band envelope well into non-universal energy scales ${\cal{O}}(t_*)$,
and because the quoted $n_c=0.6$ is far from its asymptotic strong
coupling value of $n_c=0.48$ (from Eq.(3.13) above) for the bare 
parameters specified. Further resolution of this matter is clearly required,
but we suspect that the parameter regime considered {\it e.g.\ }in
Figs.6,7 of [8]
was not sufficiently strong coupling to uncover a well developed
spectral gap.

  The scaling spectra shown in Fig.8 refer specifically to `LMAI' as detailed 
in \S 4.3, on which we focus in this paper. In Fig.9 however
we compare the resultant $f$-electron scaling spectra
$\pi\Delta_{0}D^{f}(\omega)$ with those arising from `LMAII', where 
(see \S 4.3) the polarization propagators entering the LMA self-energies
are further renormalized in terms of the host/medium propagators
$\{\tilde{\cal{G}}_{\sigma}(\om) \}$; again for the HCL with
$\ep_{c} =0.2$ and $\eta =0$. The inset shows the LMAI/II comparison out
to $\om^{\prime} = 10$, while the main figure extends to much larger
scales. And the two spectra are seen to be essentially coincident on all
$\om^{\prime}$ scales (as they ought to be if the LMA captures adequately 
the scaling spectrum).
\begin{figure}[h]
\begin{center}
\includegraphics[scale=0.32]{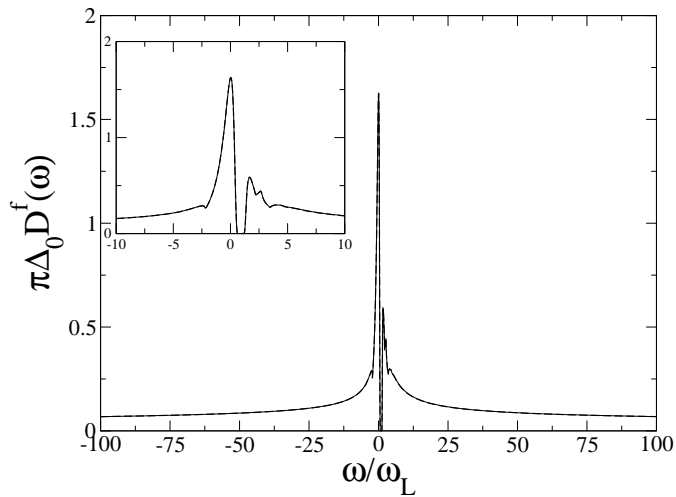}
\end{center}
\protect\caption{
Comparison of $f$-electron scaling spectra arising from LMAI
(solid line) and LMAII (dashed line), as explained in text:
$\pi\Delta_{0}D^{f}(\omega)$ {\it vs } $\om/\om_{L}$ for $\ep_{c} =0.2$
and $\eta =0$. The two levels of LMA yield essentially coincident
scaling spectra.
}
\end{figure}

 Fig.8 above illustrates that universal spectral scaling, independent
 of $U$ and $V^{2}$, arises for fixed
$\ep_{c}$ and $\eta$ which embody respectively asymmetry in the conduction
band and $f$-levels. This we find to be quite general: $D^{c}(\om)$ and/or 
$\pi\Delta_{0}D^{f}(\omega)$ exhibit scaling as an \it entire \rm function of 
$\om^{\prime} = \om/\om_{L}$ only for fixed $(\ep_{c},\eta)$,
{\it i.e.\ }distinct scaling spectra arise for different $(\ep_{c},\eta)$.
Much more subtly however, the $\ep_{c}$- and $\eta$-dependences of
the scaling spectra depend upon the $\om^{\prime}$-regimes considered,
as now explained. We begin with the simple case of low-$\om^{\prime}$. 
As pointed out in \S 3, the quasiparticle spectra Eq.(3.11) imply 
that --- in their $\om^{\prime}$-regime
of validity --- the scaling spectra should actually be independent 
of the $f$-level asymmetry $\eta$. That this is recovered within the 
LMA is seen in Fig.10 for the
HCL where, for fixed $\ep_{c} =0.3$, the $f$-electron scaling spectra 
$\pi\Delta_{0}D^{f}(\om)$ are shown for three different $f$-asymmetries
$\eta =0, 0.2,0.4$. For $|\om^{\prime}| \lesssim 1$, the regime where the
quasiparticle forms hold, the LMA scaling spectra are indeed seen to 
be independent of $\eta$; while for larger-$|\om^{\prime}|$ by contrast,
an $\eta$-dependence to the spectra is evident (and discussed further
below). Likewise, for $|\om^{\prime}| \lesssim 1$, the quasiparticle
forms Eq.(3.11) show that the scaling spectra depend explicitly on 
$\ep_{c}$, as well as on the underlying
lattice itself (embodied in the specific form for $\rho_{0}(\om)$).
\begin{figure}[h]
\begin{center}
\includegraphics[scale=0.32]{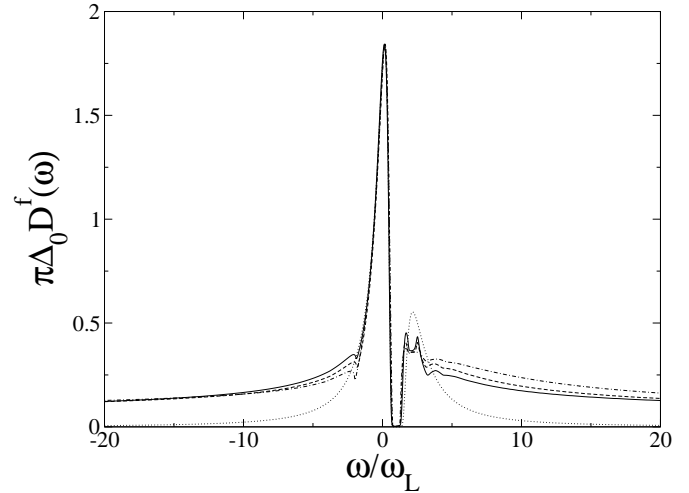}
\end{center}
\protect\caption{
Independence of the low-$|\om^{\prime}|$ scaling spectra on
the $f$-level asymmetry, $\eta$. For the HCL with fixed $\ep_{c} =0.3$,
$\pi\Delta_{0}D^{f}(\om)$ {\it vs } $\om^{\prime} =\om/\om_{L}$
is shown for $\eta =0$ (solid line), $0.2$ (dashed line) and $0.4$ (point dash).
The limiting low-$|\om^{\prime}|$ quasiparticle form is again shown for
comparison (dotted line).
}
\end{figure}

  But what of higher energies in the scaling spectra? Here as we now show
the low-$\om^{\prime}$ situation above is reversed: the high-energy
scaling behaviour of the $f$-electron spectrum is dependent on the asymmetry 
$\eta$, but 
independent of both $\ep_{c}$ and the underlying host lattice;
the latter in turn being intimately related to the emergence of effective
single-impurity physics in the high-energy scaling behaviour of the PAM.

\begin{figure}[t]
\begin{center}
\includegraphics[scale=0.32]{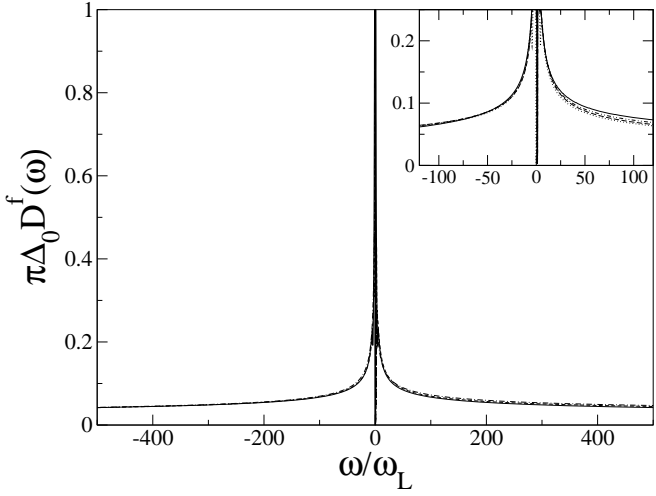}
\end{center}
\protect\caption{
$\pi\Delta_{0}D^{f}(\om)$ {\it vs } $\om^{\prime}
=\om/\om_{L}$ for the HCL up to $|\om^{\prime}| =500$, for $\eta =0$ 
and three different
$\ep_{c} =0$ (solid line), $0.3$ (dashed line) and $0.5$ (point dash); 
obtained explicitly for $U \simeq 6.6$ and $V^{2} =0.2$. 
Inset: $\pi\Delta_{0}D^{f}(\om)$ {\it vs } $\om^{\prime}$ for 
fixed $\ep_{c} =0.3$ with increasing interaction strength 
$x=\tfrac{1}{2}U|\mu| = 2.0$ (solid line), $2.5$ (double point dash) and
$3.25$ (long dash); the true scaling limit is also indicated 
(dotted line). Full discussion in text.
}
\end{figure}
To see this, Fig.11 (for the HCL) shows $\pi\Delta_{0}D^{f}(\om)$
{\it vs}  $\om^{\prime} = \om/\om_{L}$ up to $|\om^{\prime}| = 500$,
for $\eta =0$ and three different $\ep_{c}= 0, 0.3, 0.5$ (progressively
diminishing conduction band filling $n_{c}$); the particular results shown 
having been obtained 
explicitly for $U \simeq 6.6$ ($x=\tfrac{1}{2}U|\mu| = 3.25$) and $V^{2} = 0.2$.
Looking at the negative-$\om^{\prime}$ side in particular, it is clear
that the slowly decaying spectral `tails' are indeed asymptotically 
common for the different $\ep_{c}$'s. On the positive-$\om^{\prime}$ side
there might appear from the figure to be a residual weak dependence of 
the spectral tails on $\ep_{c}$. That however is simply a reflection of 
the natural fact that the 
value of $U/V^{2}$ required to reach the full asymptotic scaling spectra 
is dependent upon $\ep_{c}$. This is illustrated further in the inset to
Fig.11, which for fixed $\ep_{c} =0.3$ shows (on an expanded scale)
the evolution of the scaling spectrum with increasing interaction 
strength: $x = 2.0, 2.5$ and $3.25$. Looking at the positive $\om^{\prime}$
side one sees that the true scaling limit (dotted line) is steadily approached
upon increasing the interaction strength, but not reached until a $U$ somewhat
in excess of $\simeq 6.6$ ($x=3.25$). On the negative $\om^{\prime}$ side by
contrast, the scaling limit is already reached by $x=2.5$ ($U\simeq 5.1$) and
does not change with further increasing interaction strength. This is why
the $\ep_{c}$-independence of the spectral tails is clearly evident
only on the $\om^{\prime} <0$ side of the main figure; upon increasing
$U$ however both sides of the scaling spectra show this behaviour.

  What then is the functional form of the
large-$|\om^{\prime}|$ spectral tails? On physical grounds one expects that
on sufficiently high energy and/or temperature scales, the $f$-electrons
in the Kondo lattice regime of the PAM should be screened in an essentially 
incoherent single-impurity fashion; and thus that effective single-impurity
physics should arise in the lattice model at high energies, quite distinct from 
the effects of lattice coherence evident on low-energy scales 
$\om/\om_{L} \sim 
{\cal{O}}(1)$. For the AIM itself the spectral tails of the local
impurity scaling spectrum $D_{imp}(\omega)$ can be obtained analytically within
the LMA [27,28]; being given by
\beq
\pi\Delta_{0}D_{imp}(\om) \sim \frac{1}{2}\left(
\frac{1}{[\tfrac{4}{\pi}\ln(c|\tilde{\om}|)]^{2} + 1} +
\frac{5}{[\tfrac{4}{\pi}\ln(c|\tilde{\om}|)]^{2} +25}
\right)
\eeq
(shown explicitly for $\eta =0$)
where $\tilde{\om} = \om/\om_{K}$ with $\om_{K}$ the Kondo scale for 
the AIM discussed in \S 5.2, and $c$ a pure constant ${\cal{O}}(1)$. For
$\tilde{\omega} \gtrsim 5-10$ or so, Eq.(5.2) is known [28] to describe 
quantitatively the
spectral tails arising from NRG calculations; the exact
high energy scaling asymptote  $\pi\Delta_{0}D_{imp}(\om) \sim
3\pi^{2}/[16\ln^{2}(\om/\om_{K})]$ being recovered in particular. If effective
single-impurity behaviour arises in the PAM on energy scales encompassed 
by the $\om^{\prime}$-scaling regime, then the spectral tails thereof 
should have the same {\it scaling form } as for the AIM, {\it i.e.\ }should
be given by
\beq
\pi\Delta_{0}D^{f}(\om) \sim \frac{1}{2}\left(
\frac{1}{[\tfrac{4}{\pi}\ln(a|\om^{\prime}|)]^{2} +1} +
\frac{5}{[\tfrac{4}{\pi}\ln(a|\om^{\prime}|)]^{2} +25}
\right)
\end{equation}
with $\om^{\prime} = \om/\om_{L}$ and $a$ a constant ${\cal{O}}(1)$.
Note that such comparison requires neither a knowledge of how the 
low-energy scales for the two distinct models ($\om_{L}$ and $\om_{K}$)
depend on the bare material
parameters, nor any assumption that the Kondo scale for the AIM itself
is at all relevant to the PAM; points to which we return again in \S 5.4.
 
\begin{figure}[h]
\begin{center}
\includegraphics[scale=0.32]{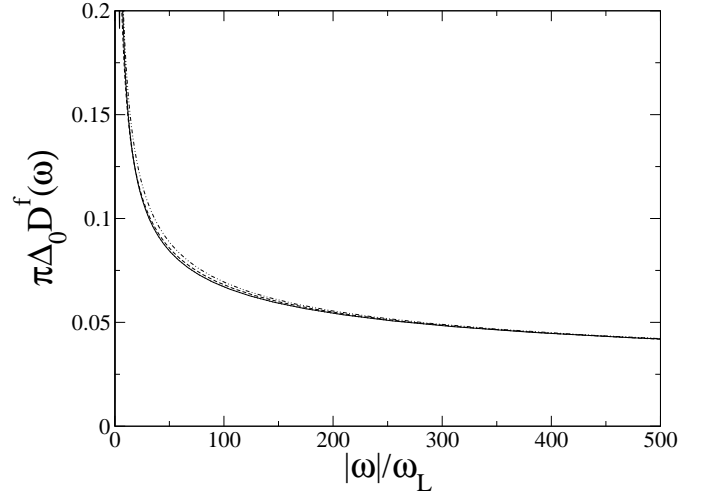}
\end{center}
\protect\caption{
Effective single-impurity physics arising in the PAM scaling spectra
at high energies, as explained in the text. The PAM $\pi\Delta_{0}D^{f}(\om)$
{\it vs } $|\om|/\om_{L}$ on an expanded vertical scale, for 
$\eta =0$ and $\ep_{c}=0$ (solid line), $0.3$ (dashed line) and $0.5$ 
(double point dash); compared to
the scaling form Eq.(5.3) (dotted line, barely distinguishable in the figure).
}
\end{figure}
Eq.(5.3) indeed describes the tail behaviour of the PAM scaling spectra,
and as such establishes the connection to effective single-impurity 
behaviour at high energies.  This is
shown explicitly in Fig.12 where, again for $\eta =0$ and $\ep_{c} =0,0.3$ and
$0.5$ (as in Fig.11), $\pi\Delta_{0}D^{f}(\om)$ is shown {\it vs}
$\omega^{\prime}$ on an expanded vertical scale, and compared to 
Eq.(5.3) (with the constant $a \simeq 0.55$ determined numerically).
That form is clearly seen to hold
asymptotically for the different $\ep_{c}$'s; and even for 
lower $\om^{\prime}$ the spectra display only a very weak dependence 
on $\ep_{c}$. Neither is it lattice dependent, the same asymptotic tail 
behaviour being found to arise whether 
the HCL or BL is considered; naturally so, for effective single-impurity
scaling physics should be independent of the `host' lattice.

\begin{figure}[h]
\begin{center}
\includegraphics[scale=0.32]{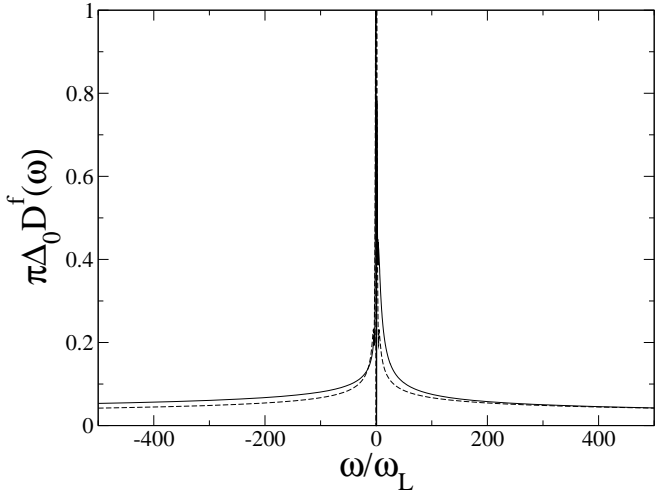}
\end{center}
\protect\caption{
$f$-electron scaling spectra $\pi\Delta_{0}D^{f}(\om)$
{\it vs } $\om^{\prime} = \om/\om_{L}$, for $\eta =0$ (solid line)
and $\eta =0.3$ (dashed line); shown explicitly for $\ep_{c} =0$ (the
scaling spectra at low-$\om^{\prime}$ depend upon $\ep_{c}$, see text).
}
\end{figure}
  While the discussion above has focused on varying $\ep_{c}$ (and hence
$n_{c}$) for symmetric $f$-levels $\eta =0$, the behaviour found is quite
general. For fixed $\eta \neq 0$ the high-energy PAM scaling spectrum 
is likewise independent of both $\ep_{c}$ and the lattice type; and 
is again found to have precisely the same scaling form as its counterpart 
for the AIM (the generalisation of Eq.(5.2) to finite-$\eta$, specifically
Eq.(5.5) of [27]).
As for the AIM [27] the resultant $f$-electron scaling spectra are now 
$\eta$-dependent, as illustrated in Fig.13 where $\pi\Delta_{0}D^{f}(\omega)$ 
is compared for $\eta =0$ and $0.3$ (and is shown specifically for 
$\ep_{c} =0$, bearing in mind that the spectra at low-$\omega^{\prime}$ 
depend on $\ep_{c}$ as discussed above). The 
$\eta$-dependence of the spectral tails is clearly evident, albeit rather
weakly so for positive $\om^{\prime}$ in particular.

  The above results capture the evolution of the
scaling spectra appropriate to the Kondo lattice regime of the PAM, from the
low-energy behaviour symptomatic of the coherent Fermi liquid state
through to the effective incoherent single-impurity physics found to arise at 
high energies --- but still in the $\om^{\prime}= \om/\om_{L}$ scaling regime.
Finally, we add that while our exclusive focus here has been on single-particle 
dynamics, the results obtained naturally have direct implications for transport 
and optical properties of heavy fermions; these will be considered
in a subsequent paper.

\subsection{Discussion: Exhaustion? }

  The discussion of the previous section brings us to Nozi\e res' issue of
`exhaustion' [37,38] and the question of how a coherent Fermi liquid state
forms in a concentrated Anderson/Kondo lattice. For a {\it single } impurity
Anderson/Kondo model, assuming [38] the only conduction  electrons eligible
to provide Kondo screening are those lying within
$\sim \om_{K}$ ($\equiv $ `$T_{K}$') of the Fermi level ($\om=0$), the number
of such is $N_{S} \sim N_{L}d^{c}_{0}(0)\om_{K}$; with $N_{L}$ the total number of lattice
sites (and $d^{c}_{0}(\om) \equiv \rho_{0}(\om -\ep_{c})$ 
the free conduction band dos, normalised to
unity). In the strong coupling Kondo regime, $d^{c}_{0}(0)\om_{K}$
is of course exponentially small; but $N_{S}$, the number of available 
screening electrons per the single impurity spin, obviously remains 
macroscopically large. 
That situation changes drastically in the concentrated Anderson/Kondo lattice.
Now there are $N_{L}$ spins ($f$-electrons) to screen; so the number of 
electrons {\it per $f$-spin } available to provide Kondo screening is 
$N_{S}/N_{L} \sim d^{c}_{0}(0)\om_{K}$ --- itself exponentially small.
That raises the issue of `exhaustion' [37,38]: how so few screening electrons 
lead to the formation of a coherent Fermi liquid ground state. Nozi\e res' 
argument [38] is that this effectively arises through a two-stage process 
with decreasing energy/temperature scale.
Neglecting the RKKY interaction, on high energy/temperature scales the 
local $f$-spins are first Kondo screened in an essentially incoherent,
single-impurity fashion; while with further decreasing energy this 
effective single-impurity regime crosses over into lattice coherent 
behaviour through collective screening/isotropization
of the $f$ spins. Two underlying scales are then argued to emerge: a 
high energy single-impurity Kondo scale $\om_{K}$ corresponding to the 
incoherent effective single-impurity physics; and a second, lower 
lattice scale $\om_{L}$ ($\equiv$ `$T_{c}$')
signifying the onset of lattice coherence. Nozi\e res has provided intuitive
arguments [38] to suggest that, at most, $\om_{L} \sim
d^{c}_{0}(0)[\om_{K}]^{2}$; which, since $\om_{K}$ itself is exponentially
small, means that $\om_{L}$ and $\om_{K}$ are radically distinct in 
scaling terms (as elaborated below).
Further, as noted by Pruschke {\it et. al.\ }[8], Nozi\e res' phenomenological
arguments are not in fact particular to low conduction band filling $n_{c}$,
and if correct imply two-scale exhaustion physics should be the generic 
situation.

  Much work has since ensued on the question of exhaustion [8,11,12,17,23] via 
DMFT studies of the PAM and/or KLM. Regarding scales {\it per se} there
appears now to be an approaching consensus [8,23] that 
$\om_{L} \propto [\om_{K}]^{2}$ does not arise; but rather that 
$\om_{L} \propto \om_{K}$, with a proportionality dependent 
on the conduction band filling $n_{c}$ (or equivalently $\ep_{c}$):
$\om_{L}/\om_{K} = F(n_{c})$, with which the present work concurs as in \S 5.2.
That granted however, it has nonetheless still been suggested {\it e.g.\ }in
[8,23] that away from half-filling where $F(n_{c}) <1$, a two-scale picture 
arises.  This underlies the qualitative notion of `protracted screening' 
from Quantum Monte Carlo/Maximum Entropy Method studies [11,12], where the
existence of two scales is inferred
because the thermal evolution of dynamics is slower (or `protracted') for 
the PAM than the AIM;  and it arises likewise in the large-$N$ mean-field 
study of [23] (where the emphasis is on low conduction band filling). It 
is also suggested in the NRG study of [8] (\S V of [8]), although we add
that the NRG results themselves were not argued in [8] to provide evidence
for relevance of the single-impurity scale $\omega_{K}$ to dynamics of 
the PAM itself.

  As is evident from the results of the previous section, we dissent from
the view of a two-scale picture (without disagreeing with specific results 
obtained {\it e.g.\ }in [8,23]). That deserves a careful explanation. The 
first point to make here is that if {\it if } two distinct scales were 
relevant to the PAM/KLM, in particular to its behaviour all the way from 
the coherent Fermi liquid through to effective single-impurity physics, then 
there should be two distinct {\it scaling regimes } of the model. 
But that could only be ascertained by investigation of the scaling
behaviour of physical properties (in the relevant strongly correlated
regime); which has not hitherto been considered.

   Let us then suppose that two distinct 
scales arise. The lower will be the coherence scale
$\om_{L}$, and (without prejudice as to its origin) call the higher
energy scale $\om_{H}$. As a function of the bare model parameters it is
taken as read that both scales vanish asymptotically in strong coupling 
({\it i.e.\ }become exponentially small, well separated from non-universal 
scales).  Now suppose the ratio 
$\om_{H}/\om_{L} \rightarrow \infty$ 
in strong coupling (which would be the case {\it e.g.\ }in the Nozi\e res 
exhaustion scenario [38] where $\om_{H}$ corresponds to the AIM scale 
$\om_{K}$ and $\om_{L} \propto [\om_{K}]^{2}$). In that case, scaling 
of physical properties in terms of $\om^{\prime} = \om/\om_{L}$ (or 
$T/\om_{L}$) would project out to infinity energies on the scale of 
$\om_{H}$ (as well as the usual irrelevant non-universal scales); this 
is the `$\om/\om_{L}$' scaling regime.
By contrast, scaling in terms of $\om^{\prime\prime} = \om/\om_{H}$
would project out both non-universal scales (to infinity as usual) 
{\it and } all
energies on the scale of $\om_{L}$ or finite multiples thereof (to zero);
this is the second, `$\om/\om_{H}$' scaling regime. The coherent Fermi
liquid will of course be encapsulated in the former, $\om/\om_{L}$ scaling
regime. Regarding the crossover to effective single-impurity behaviour in
the PAM/KLM, two obvious possibilities arise. (a) That this crossover is set
by the $\om_{H}$ scale (as in the Nozi\e res picture). It will then  
arise only in the second scaling regime; effective single impurity behaviour
will not thus be evident as a function of $\om/\om_{L}$. (b) That
the crossover occurs in the $\om/\om_{L}$ scaling regime. In that case
the second putative scaling regime is irrelevant --- at least to 
the central issue of understanding the evolution of the PAM/KLM from the 
coherent Fermi liquid through to effective single impurity behaviour.

  The behaviour hypothesised above reflects a genuine two-scale description.
But if by contrast the ratio $\om_{H}/\om_{L}$ is a constant in strong 
coupling then the scales $\om_{L}$ and $\om_{H}$ are equivalent,
differing only by the constant but fundamentally equivalent in scaling terms.
In this case there is no essential distinction between $\om_{L}$ and
$\om_{H}$, and obviously only one scaling regime. 
Such behaviour is of course well known to arise {\it e.g.\ }in the Anderson
impurity model, where the `Kondo scale' appears in many different but 
equivalent guises: {\it e.g.\ }[2] the usual Kondo temperature
$T_{K}$ obtained from the impurity susceptibility, the half-width at half 
maximum of the Kondo/Abrikosov-Suhl resonance in the single-particle 
spectrum, or $\Delta_{0}Z_{imp}$ with $Z_{imp}$ the impurity quasiparticle 
weight. Each is proportional
to the other; all are manifestations of the single underlying Kondo scale.

  Within the present theory the results of \S 5.3 show that 
a one-scale picture arises: the evolution from a coherent Fermi liquid 
to effective single-impurity physics arises clearly when the scaling behaviour
of dynamics is considered as a function of $\om/\om_{L}$ (and we find
no evidence for a `higher' scaling regime). Neither is such behaviour
confined to single-particle dynamics. As will be shown in a subsequent paper
the resistivity $\rho(T)$, including its crossover to effective single-impurity
behaviour, likewise exhibits one-parameter universal scaling in terms of 
$T/\om_{L}$. In parallel to the above comments on the AIM, that does 
{\it not } of course preclude the coherence scale appearing in different but
equivalent guises. 
For example the peak maximum in the resistivity, often chosen as a 
measure of the low-energy scale, is not identically $\om_{L}$; but the two
are simply proportional, indicative of one-parameter scaling. 

  One further point should be noted here.
The connection to effective single-impurity physics in
the PAM/KLM established in \S 5.3 arises from comparison of the
{\it scaling forms} of the spectra for the PAM/KLM and the AIM, 
{\it i.e.\ }as functions of
$\om/\om_{L}$
and $\om/\om_{K}$ respectively. That does not require {\it any }
knowledge of how the separate scales for the two models, $\om_{L}$ and 
$\om_{K}$, themselves depend on the bare material parameters.
Neither does it require any assumption that the Kondo scale $\om_{K}$ 
for the single-impurity model {\it itself} is directly relevant 
to the PAM/KLM (the two are after all different models).  It is nonetheless
the case that, from specific study (\S 5.2)
of how the respective scales for the two models depend upon the bare
parameters, we find $\omega_{L}/\omega_{K} = F(n_{c})$
in agreement with previous work [8,23]; and while that behaviour is 
neither required 
nor relevant in establishing via scaling the connection to effective 
single-impurity physics in the PAM/KLM, it does
mean that the two scales are fundamentally {\it equivalent} if used formally as 
scaling variables, {\it i.e.\ }one may 
choose equivalently to use $\om_{K}$ to scale the PAM spectra. Hence, 
while we concur with the basic results of [8,23] for the relation between
$\om_{K}$ and $\om_{L}$, we naturally disagree with the view 
that a two-scale picture arises.

  We also note parenthetically that the single-scale picture found here 
simply obviates an apparent conundrum raised in [8], namely how one 
rationalises the regime $n_{c} \approx 1$ where $\om_{L}/\om_{K} > 1$ --- the
lattice coherence scale is now larger than the single-impurity Kondo scale.
As pointed out in [8] this presents a self-evident problem for 
interpretations based on the assumption 
that the AIM $\om_{K}$ is the relevant scale for effective single-impurity
physics in the PAM/KLM, while $\om_{L}$ sets the scale for lattice coherence.
From the viewpoint of the present work however, this is obviously not 
an issue; effective single-impurity physics in the PAM/KLM arises as 
naturally for $n_{c} \rightarrow 1$ (see also [34,35]) as it does for lower 
conduction band fillings $n_{c}$ where $\om_{L}/\om_{K} <1$.
\begin{figure*}[t]
\begin{center}
\includegraphics*[scale=0.6]{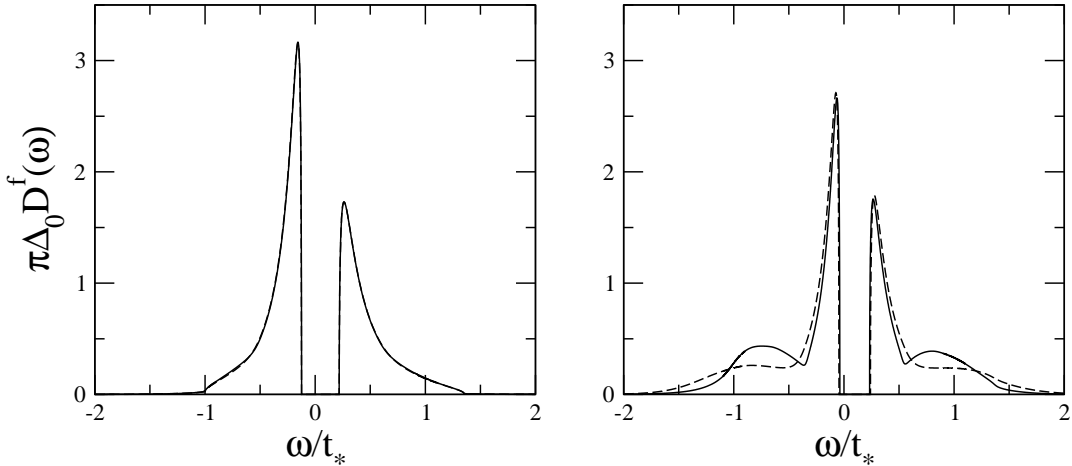}
\end{center}
\protect\caption{
Weak coupling LMA $f$-electron spectrum (solid
line) {\it vs } $\om/t_{*}$ for the Bethe lattice, with
$U=0.75$ (right panel) and  $U=0.25$ (left panel); and
for $\ep_c=0.2$, $\eta =0$ and $V^2=0$.
Corresponding  results from second order perturbation
theory in $U$ are also shown (dashed lines); for the
lower-$U$ shown this is indistinguishable from the
LMA spectrum.
}
\end{figure*}

  The obvious conclusion from the preceding discussion is that we find 
no compelling evidence for two-scale exhaustion.  A sceptic can naturally 
argue that since the present theory is approximate it is open to doubt. 
That is of course true, as it is for any theory. But the evidence here 
certainly points away from exhaustion and, should further support for the 
idea be forthcoming, it will in our view require convincing scaling 
arguments to be established. 

\section{Concluding remarks}

We have developed in this paper a local moment approach to single-particle
dynamics of the periodic Anderson model within the framework of dynamical
mean-field theory, for the generic asymmetric case relevant to heavy 
fermion metals. For obvious physical reasons our primary interest has been 
the strongly correlated Kondo lattice regime, of essentially localized 
spins $n_f \rightarrow 1$ but with general conduction band filling $n_c$,
which the intrinsically non-perturbative nature of the LMA renders readily
accessible. The exponentially small lattice coherence scale $\om_L$ 
inherent to the Kondo lattice regime leads in particular to a clean 
separation of low-energy (`universal') and high-energy scales, and
hence to universal scaling behaviour of dynamics. This has been a central 
focus of the present work and a rich description of scaling spectra
results, spanning all $\om/\om_L$-scales. With increasing $\om^\prime=
\om/\om_L$, dynamics are found to cross over from the low-energy
quasiparticle behaviour symptomatic of the coherent Fermi liquid state
to essentially incoherent single-impurity Anderson/Kondo scaling physics
at high-$\om^\prime$. The former, low-$\om^\prime$ behaviour depends
naturally on both the conduction band filling and underlying lattice `type'.
The latter by contrast depends on neither, consistent with one's
physical expectation of effective single-impurity physics; and the crossover
from coherent Fermi liquid to effective single-impurity behaviour in
the PAM, established as it is by scaling, neither presumes nor
requires any particular relation between the PAM coherence scale, $\om_L$, 
and the Kondo scale $\om_K$ for the corresponding `real' AIM
arising when only a single $f$-level is coupled to the conduction
band.

While our almost exclusive emphasis has been on the strongly correlated 
regime we add that, as for the Anderson impurity models considered
hitherto [26,27], all interaction strengths are nonetheless encompassed
by the LMA including simple perturbative behaviour in weak coupling
-- an illustration of the latter being given in Fig.14 where, for
$U=0.75$ and $0.25$ (with $\ep_c=0.2, \eta=0$ and $V^2=0.2$) LMA
results for the $f$-electron spectrum $\pi\Delta_0D^f(\om)$ {\it vs}
$\om/t_*$ are compared to those arising from second order
perturbation theory (PT) in the interaction $U$: with decreasing 
interaction the LMA spectrum clearly reduces to that arising 
from PT, being indistinguishable from it for the lower $U$ shown.

The essential criteria for a successful description of the PAM thus
appear to be met by the LMA; all energy scales, and interaction
strengths from weak to strong coupling, being handled in a unified theoretical
framework. Further (as mentioned in \S 1), a description of the single-particle
dynamics considered here is a prerequisite to determining transport
and optical properties of heavy fermions; which subject, including
explicit comparison to experiment, will be considered in a forthcoming
paper.

\begin{acknowledgement}
We express our thanks to the EPSRC for supporting
this research.
\end{acknowledgement}

\end{document}